\documentclass[letterpaper,twocolumn,10pt]{article}  

\usepackage{usenix2019_v3,epsfig,endnotes}
\usepackage[normalem]{ulem}
\usepackage{makecell} 
\usepackage{algpseudocode}

\ifx\noeditingmarks\undefined
    
\else
    
\fi

\ifx\noeditingmarks\undefined
   \newcommand{\pgwrapper}[3]{\begingroup \color{#1} #2: #3 \endgroup}
   \newcommand{\pgwrapperb}[1]{\textbf{#1}}
   \newcommand{\dangerwrapper}[1]{{\color{red}#1}}
\else
   \newcommand{\pgwrapperb}[1]{}
   \newcommand{\pgwrapper}[3]{}
   \newcommand{\dangerwrapper}[1]{}
\fi

\newcommand{\ex}[0]{\xrightarrow{\text{\raisebox{-.3ex}[0ex][-.5ex]{exe}}}}{}

\newcommand{\rt}[0]{\xrightarrow{\text{\raisebox{-.3ex}[0ex][-.5ex]{rto}}}}{}
\newcommand{\EX}{\overset{\text{\raisebox{-.3ex}[0ex][-.5ex]{\textbf{exe}}}}{\longmapsto}}

\usepackage{tikz}
\newcommand*\circled[1]{\tikz[baseline=(char.base)]{
            \node[shape=circle,draw,inner sep=.75pt, line width=.85pt, font=\footnotesize\bfseries] (char) {#1};}}

\usepackage{textcomp}

\newcommand*\mycircled[1]{\tikz[baseline=(char.base)]{
            \node[shape=circle, draw, inner sep=0.5pt, fill=black, line width=.85pt, font=\footnotesize] (char) {\textcolor{white}{#1}};}}

\usepackage{pifont}

\newcommand{\AUTHORS}{Authors}

\newcommand{\TITLE}{NCC: Natural Concurrency Control for Strictly Serializable Datastores by Avoiding the Timestamp-Inversion Pitfall\\(Extended Version)\\\large Technical Report arXiv:2305.14270\thanks{This technical report is an extended version of the paper under the same title that appeared in OSDI 2023~\cite{lu-osdi23-ncc}.}}

\newcommand{\KEYWORDS}{Put your keywords here}
\newcommand{\CONFERENCE}{Somewhere}
\newcommand{\PAGENUMBERS}{yes}      
\newcommand{\COLOR}{yes}
\newcommand{\onlyAbstract}{no}
\newcommand{\newpagesections}{no}
\newcommand{\makePasteable}{no}  

\usepackage{ifthen}
\ifthenelse{\equal{\PAGENUMBERS}{yes}}{%
\usepackage[nohead,
            left=.75in,right=.75in,top=1in,
            footskip=0.5in,bottom=1in,     
            columnsep=0.33in
            ]{geometry}
}{%
\usepackage[noheadfoot,left=.75in,right=.75in,top=1in,
            footskip=0.5in,bottom=1in,
            columnsep=0.33in
	    ]{geometry}
}

\ifthenelse{\equal{\newpagesections}{yes}}{%
\newcommand{\nps}{\realnewpage}
}{%
\newcommand{\nps}{}
}

\usepackage[font=bf]{caption}
\usepackage[font=bf]{subcaption}
\usepackage[font=bf,aboveskip=0pt]{caption} 
\usepackage[compact]{titlesec}

\titlespacing{\paragraph}{0pt}{*1.0}{*1}      
\usepackage{etoolbox}
\makeatletter
\patchcmd{\ttlh@hang}{\parindent\z@}{\parindent\z@\leavevmode}{}{}
\patchcmd{\ttlh@hang}{\noindent}{}{}{}
\makeatother

\usepackage{enumitem}
\setlist{itemsep=1.5pt,parsep=1.5pt}             

\usepackage{fancyhdr}

\ifthenelse{\equal{\PAGENUMBERS}{yes}}{%
  \pagestyle{plain}
}{%
  \pagestyle{empty}
}

\usepackage{booktabs}
\usepackage{color}
\usepackage{colortbl}
\usepackage{float}                           

\usepackage{mathptmx}                        
\usepackage{courier}
\usepackage[scaled=0.92]{helvet}
\usepackage{centernot}                       

\ifthenelse{\equal{\COLOR}{yes}}{%
  \usepackage{hyperref}
}{%
  \usepackage[pdfborder={0 0 0}]{hyperref}
}

\hypersetup{%
pdfauthor = {\AUTHORS},
pdftitle = {\TITLE},
pdfsubject = {\CONFERENCE},
pdfkeywords = {\KEYWORDS},
bookmarksopen = {true}
}

\definecolor{placeholderbg}{rgb}{0.85,0.85,0.85}

\ifthenelse{\equal{\makePasteable}{yes}}{%
  \hyphenpenalty 10000                                                                  
  \exhyphenpenalty 10000                                                                
  \setlength{\paperheight}{10740pt}                                                      
  \setlength{\textheight}{10740pt}
  \onecolumn
}{%
}

\usepackage{soul}
\usepackage{xspace}
\usepackage{fancyvrb}
\usepackage{multirow}
\usepackage{amssymb} 
\usepackage{titling}
\usepackage[utf8]{inputenc}
\usepackage[english]{babel}
\usepackage{amsthm}
\usepackage{textcomp} 
\usepackage{balance} 
\usepackage{bm} 
\usepackage{upgreek} 
\usepackage[shortcuts]{extdash}
\usepackage{amsthm}
\usepackage{mathtools} 

\usepackage{pdflscape}  

\makeatletter
\newtheorem*{rep@theorem}{\rep@title}
\newcommand{\newreptheorem}[2]{%
\newenvironment{rep#1}[1]{%
 \def\rep@title{#2 \ref{##1}}%
 \begin{rep@theorem}}%
 {\end{rep@theorem}}}
\makeatother

\newtheorem{corollary}{Corollary}
\newtheorem{lemma}[corollary]{Lemma}
\newreptheorem{lemma}{Lemma}
\newcommand{\proofstart}{\noindent \textit{Proof.} }
\newcommand{\proofend}{{\hfill{}\footnotesize$\blacksquare$}}

\usepackage[linesnumbered,titlenumbered,ruled,vlined,noresetcount,algosection]{algorithm2e}
\definecolor{cb-brown-light}{RGB}{216, 179, 101}
\definecolor{cb-blue-light}{RGB}{182, 219, 255}

\newcommand{\sysname}{Natural Concurrency Control\xspace} 
\newcommand{\sstx}{NCC\xspace}  
\newcommand{\sstxrw}{NCC-RW\xspace}
\newcommand{\tplww}{d2PL-wound-wait\xspace} 
\newcommand{\tplnw}{d2PL-no-wait\xspace}

\newcommand{\inversion}{timestamp-inversion}

\newcommand{\tld}{$\sim$}

\newcommand{\reflem}[1]{Lemma~\ref{thm:#1}}

\theoremstyle{definition}
\newtheorem{definition}{Definition}[subsection]

\newcommand{\INV}[1]{{ {INV}(#1)}}
\newcommand{\RESP}[1]{{ {RESP}(#1)}}

\newcommand{\ROM}[1]{\MakeUppercase{\romannumeral #1}}

\usepackage[absolute]{textpos}
\usepackage[yyyymmdd]{datetime}

\newfloat{acmcr}{b}{acmcr}

\definecolor{eblue}{RGB}{0, 0, 139}
\definecolor{mblue}{RGB}{0, 147, 175}
\definecolor{rgreen}{RGB}{0, 112, 60}
\definecolor{worange}{RGB}{245, 128, 37}
\definecolor{jred}{RGB}{255, 0, 0}
\definecolor{hpurple}{rgb}{0.6, 0.4, 0.8}
\definecolor{bpurple}{rgb}{0.54, 0.17, 0.89}
\definecolor{seagreen}{rgb}{0.18, 0.55, 0.34}
\definecolor{sinopia}{rgb}{0.8, 0.25, 0.04}

\newcommand\KW[1]{\textbf{#1}} 
\newcommand\CM[1]{\emph{#1}}   
\definecolor{owblue}{RGB}{0, 0, 96}
\definecolor{frgreen}{RGB}{0, 96, 48}

\newcommand{\sg}{safeguard\xspace}
\newcommand{\Sg}{Safeguard\xspace}

\newcommand{\sr}{smart retry\xspace}
\newcommand{\Sr}{Smart retry\xspace}

\usepackage{cleveref}
\usepackage{chngcntr}
\newcounter{invariants}
\crefname{invariant}{Invariant}{Invariants}

\newcommand{\invariant}[2]{\textbf{Invariant \refstepcounter{invariants}\label[invariant]{#1}\theinvariants:} #2}

\makeatletter
\newcommand{\setword}[2]{%
  \phantomsection
  #1\def\@currentlabel{\unexpanded{#1}}\label{#2}%
}
\makeatother

\date{\vspace{-3ex}}
\title{\textbf{\TITLE}\vspace{-1.5ex}}
\author{Haonan Lu{$^\star$}, Shuai Mu{$^\dag$}, Siddhartha Sen{$^\ddagger$}, Wyatt Lloyd{$^\diamond$}\\
\textit{{$^\star$}University at Buffalo}, \textit{{$^\dag$}Stony Brook University},\\ \textit{{$^\ddagger$}Microsoft Research}, \textit{{$^\diamond$}Princeton University}\\
}

\begin{document}

\maketitle

\begin{abstract}
Strictly serializable datastores 
greatly simplify 
application development. 
However, existing techniques pay unnecessary costs 
for naturally consistent transactions, which arrive at servers in an order 
that is already strictly serializable. 
We exploit this natural arrival order 
by executing transactions with minimal costs while optimistically assuming they are naturally consistent, 
and then leverage a timestamp-based technique to efficiently verify if the 
execution is indeed consistent. 
In the process of this design,  
we identify a fundamental pitfall in relying on timestamps to provide 
strict serializability
and name it the \inversion{} pitfall. 
We show that timestamp inversion has affected several existing systems. 

We present \sysname{} (\sstx{}), 
a new concurrency control technique that guarantees strict serializability 
and ensures minimal costs---i.e., one-round latency, lock-free, and non-blocking execution---in the common case by leveraging natural consistency. 
\sstx{} is enabled by three components: 
non-blocking execution, decoupled response management, 
and timestamp-based consistency checking. 
\sstx{} avoids the \inversion{} pitfall 
with response timing control 
and proposes two optimization techniques, 
asynchrony-aware timestamps and 
\sr, to reduce false aborts. 
Moreover, \sstx{} designs a specialized protocol for read-only transactions, 
which is the first to achieve optimal best-case 
performance while guaranteeing strict serializability without relying on synchronized clocks. 
Our evaluation shows \sstx outperforms 
state-of-the-art strictly serializable solutions by an order of magnitude 
on many workloads.
\end{abstract}

\ifthenelse{\equal{\onlyAbstract}{no}}{%
\nps\section{Introduction}
\label{sec:intro}
Strictly serializable datastores 
have been advocated by much recent work~\cite{Corbett:osdi2012, faunadb, li2017eris, Mu:osdi2016, 
shamis2019fast, yan2018carousel, fan2019ocean} because 
they provide the powerful abstraction of programming in a single-threaded, 
transactionally isolated environment, 
which greatly simplifies application development and prevents consistency anomalies~\cite{google18why}. 
However, 
only a few concurrency control techniques provide strict serializability
and 
they are expensive. 

Common techniques include 
distributed optimistic concurrency control (dOCC), 
distributed two-phase locking (d2PL), and transaction reordering (TR).
They incur high overheads 
which manifest in extra rounds of messages, distributed lock management, 
blocking, 
and excessive aborts.
The validation round in dOCC, required lock management in d2PL, blocking during 
the exchange of ordering information in TR, and  
aborts due to conflicts in dOCC and d2PL are examples of these four overheads, respectively.
These costs are paid to enforce the two requirements of strict serializability:
(1) ensuring there is a total order 
by avoiding interleaving transactions,
and
(2) ensuring the \textit{real-time ordering} 
i.e., 
later-issued transactions take effect after previously-finished ones. 
However, we find these costs are unnecessary for many \textit{datacenter workloads} 
where transactions are executed within a datacenter 
and then replicated within or across datacenters. 

Many datacenter transactions do not interleave: 
e.g., many of them are dominated by reads~\cite{Corbett:osdi2012}, and the interleaving of reads returning the same value does not affect correctness. 
Many of them are short~\cite{hstore_one_shot, Mu:osdi2016, Lu:osdi2016, Zhang:sosp2015, kraska2013mdcc, tu2013speedy}, and short lifetimes reduce the likelihood of interleaving. 
Advances in datacenter networking also reduce variance in 
delivery times of concurrent requests~\cite{grosvenor2015queues, dpdk, infiniband},  resulting in less interleaving. 

In addition, many datacenter transactions arrive at servers in an order that trivially satisfies 
their real-time order requirement. 
That is, a transaction arrives at all participant servers after all previously committed transactions.

Because many transactions do not interleave and their arrival order satisfies 
the real-time order constraints, 
intuitively, 
simply executing their requests in the order servers receive them 
(i.e., treating them as if they were non-transactional simple operations) 
will naturally satisfy strict serializability. 
We call these transactions \textit{naturally consistent}. 

Ideally, naturally consistent transactions 
can be safely executed without any concurrency control, incurring zero costs. 
However, 
existing techniques pay unnecessary overheads. 
For instance, 
dOCC still requires extra rounds of messages for validation, 
d2PL still acquires locks, 
and  
TR still blocks transactions to exchange ordering information, 
even if validation always succeeds, 
locks are always available, 
and nothing needs to be reordered. 
Therefore, this paper strives to make naturally consistent transactions as cheap as possible.

In this paper, we present \sysname{} (\sstx{}), 
a new concurrency control technique that guarantees strict serializability 
and ensures minimal costs---i.e., one-round latency, lock-free, and non-blocking execution---in the common case. 
\sstx{}'s design insight is to execute naturally consistent transactions 
in the order they arrive, as if they were non-transactional operations, 
while guaranteeing correctness without  
interfering with transaction execution. 

\sstx{} is enabled by three components. 
\textit{Non-blocking execution} ensures that servers execute transactions in a way that is 
similar to executing non-transactional operations. 
\textit{Decoupled response management} separates the execution of requests from the sending of their responses, ensuring that only correct  
results are returned.  
\textit{Timestamp-based consistency checking} uses timestamps to 
verify transactions' results, without interfering with execution.

While designing the consistency-checking component, 
we identified a correctness pitfall in timestamp-based, strictly serializable techniques. 
Specifically, 
these techniques sometimes fail to guard against an execution order that is 
total but incorrectly inverts the real-time ordering between transactions, 
thus violating strict serializability. 
We call this the \textit{\inversion{}} pitfall. 
Timestamp inversion is subtle because it can happen only if a transaction interleaves 
with a set of \emph{non}-conflicting transactions that have real-time order relationships. 
The pitfall is fundamental as we find it affects multiple prior systems 
(TAPIR~\cite{Zhang:sosp2015} 
and DrTM~\cite{Wei:sosp2015}), which, as a result, do not provide strict serializability 
as claimed. 

\sstx{} handles timestamp inversion through  
response timing control (RTC), 
an integral part of decoupled response management,
without interfering  
with non-blocking execution or relying on synchronized clocks. 
\sstx{} proposes two timestamp optimization techniques, asynchrony-aware timestamps and \sr{}, to reduce false aborts.  
Moreover, 
\sstx{} designs a specialized protocol for read-only transactions, 
which, to the best of our knowledge, is the first to achieve optimal 
performance~\cite{Lu:osdi2016} in the best case while ensuring strict serializability, without relying on synchronized clocks. 

We compare \sstx with common strictly serializable techniques: 
dOCC, d2PL, and TR,  
and two serializable protocols, TAPIR~\cite{Zhang:sosp2015} and MVTO~\cite{reed1983implementing}.
We use three workloads: 
Google-F1, Facebook-TAO, and TPC-C (\S\ref{sec:eval}). 
The Google-F1 and Facebook-TAO workloads synthesize production-like workloads for Google's Spanner~\cite{Corbett:osdi2012,shute2013f1} 
and 
Facebook's TAO~\cite{Bronson:atc2013}, 
respectively.
Both workloads are read-dominated. 
TPC-C~\cite{tpcc} consists of few-shot transactions that are write-intensive. 
We further explore the workload space by varying the write fractions in Google-F1. 
\sstx significantly outperforms 
dOCC, d2PL, and TR 
with 
$2$--$10\times$ lower latency 
and $2$--$20\times$ higher throughput.
\sstx outperforms TAPIR with $2\times$ higher throughput and $2\times$ lower latency, and 
closely matches the performance of MVTO. 

In summary, this work makes the following contributions:
\begin{itemize}[leftmargin=*] 

\item Identifies timestamp inversion, a fundamental correctness pitfall 
in timestamp-based, strictly serializable concurrency control techniques. 

\item Proposes \sstx, 
a new concurrency control technique that provides strict serializability 
and achieves minimal overhead in the common case by exploiting 
natural consistency in datacenter workloads. 

\item A strictly serializable read-only protocol with optimal best-case performance that does not rely on synchronized clocks.

\item An implementation and evaluation that shows \sstx  
outperforms existing strictly serializable systems by 
an order of magnitude 
and closely matches the performance of  
systems that provide weaker consistency. 
\end{itemize}

\nps\section{Background}
\label{sec:background}
This section provides the necessary background on 
transactional datastores, 
strict serializability, 
and general techniques for providing strict serializability. 

\subsection{Transactional Datastores}
\label{subsec:bg}
Transactional datastores are the back-end workhorse of many web applications. 
They typically consist of two types of machines.
Front-end \textit{client} machines receive users' requests, e.g., managing a web page, 
and execute these requests on behalf of users by issuing transactions to the storage \textit{servers} that store the data. 
Servers are fault-tolerant, 
e.g., the system state is made persistent on disks and 
replicated via replicated state machines (RSM), like Paxos~\cite{lamport2001paxos}.  

Transactions are managed by coordinators, 
which can be co-located either with a server or the client. 
This paper adopts the latter approach to avoid the delays caused by shipping 
the transaction from the client to a server, 
while explicitly handling client failures. 
The coordinator issues read/write operations to relevant servers, 
called \emph{participants}, following the transaction's logic, 
which can be \emph{one-shot}, i.e., it knows a priori which data to read/write and can send all requests in one step, 
or \emph{multi-shot}, i.e., it takes multiple steps as the data read in one step determines which data to read/write in later steps. 
The system executes transactions following a concurrency control protocol, which ensures that 
transactions appear to take effect in an order that satisfies the system's consistency requirements.
The stronger the consistency provided by the system, the easier it is to develop correct applications. 

\begin{figure*}[t]
\centering
\begin{subfigure}[b]{0.3\linewidth}
  \centering
  \includegraphics[width=1.0\linewidth]{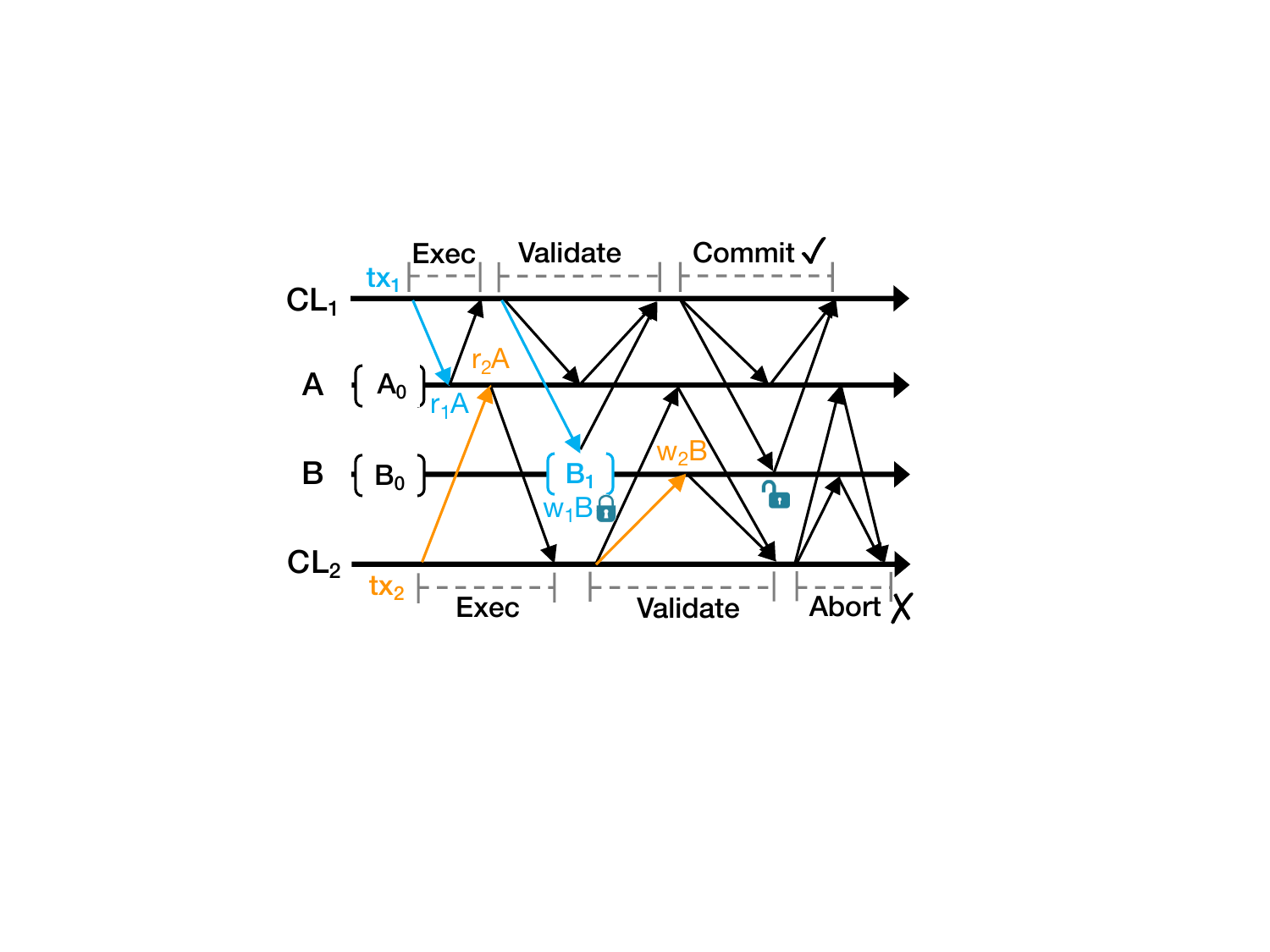}
  \caption{dOCC falsely aborts $\textit{tx}_2$}
  \label{fig:bg-1}
\end{subfigure}
~~~~~~~~
\begin{subfigure}[b]{0.3\linewidth}
  \centering
  \includegraphics[width=1.0\linewidth,page=2]{figs/bg.pdf}
  \caption{Timestamp refinement examples}
  \label{fig:bg-2}
\end{subfigure}
~~~~~~~~
\begin{subfigure}[b]{0.3\linewidth}
  \centering
  \includegraphics[width=1.0\linewidth,page=3]{figs/bg.pdf}
  \caption{\sstx{} commits both $\textit{tx}_1$ and $\textit{tx}_2$}
  \label{fig:bg-3}
\end{subfigure}
\caption{$\textit{tx}_1$ and $\textit{tx}_2$ 
are naturally consistent. 
dOCC incurs unnecessary validation costs, and $\textit{tx}_2$ could be falsely aborted due to lock unavailability. 
\sstx can commit both transactions with timestamp pre-assignment, refinement, and 
the \sg{} check (denoted by SG). These techniques are detailed in Section~\ref{subsubsec:lowcost}. 
Each version in \sstx{} has a ($t_w$, $t_r$) pair which is included in server responses. 
RTC means response timing control, detailed in Section~\ref{subsec:avoid-tip}.}
\label{fig:bg}
\end{figure*}

\subsection{Strict Serializability}
\label{subsec:ss}
\emph{Strict serializability}~\cite{herlihy90linearizability, Papadimitriou79serializability}, also known as external consistency~\cite{gifford1981information}, is often considered the strongest 
consistency model. 
It requires that 
(1) there exists a \textit{total order} of transactions, and 
(2) the total order must respect the \textit{real-time order}, which means 
if transaction $\textit{tx}_1$ ends 
before $\textit{tx}_2$ starts,
then $\textit{tx}_1$ must appear before $\textit{tx}_2$ in the total order.
As a result, transactions appear to take effect one at a time in the order the system receives them. 

\paragraph{Formal definition.} 
We use Real-time Serialization Graphs (RSG)~\cite{adya1999weak} to formalize the total order and real-time order requirements. 
An RSG is a directed graph that captures the order in which transactions take effect. 
Specifically, two requests from different transactions have an \textit{execution edge} $\textit{req}_1 \ex \textit{req}_2$ if any of the following happens: 
$\textit{req}_1$ creates some data version $v_i$ and $\textit{req}_2$ reads $v_i$; 
$\textit{req}_1$ reads some data version $v_j$ and $\textit{req}_2$ creates $v$'s next version that is after $v_j$; 
or $\textit{req}_1$ creates some data version $v_k$ and $\textit{req}_2$ creates $v$'s next version that is after $v_k$. 
Two transactions have an execution edge $\textit{tx}_1 \ex \textit{tx}_2$ if there exist $\textit{req}_1$ and $\textit{req}_2$ from $\textit{tx}_1$ and $\textit{tx}_2$, respectively, such that $\textit{req}_1 \ex \textit{req}_2$. 
A chain of execution edges constructs a directed path between two transactions (requests), 
denoted by $\textit{tx}_1 \EX \textit{tx}_2$ ($\textit{req}_1 \EX \textit{req}_2$), 
meaning that $\textit{tx}_1$ ($\textit{req}_1$) affects $\textit{tx}_2$ ($\textit{req}_2$) 
through some intermediary transactions (requests). 
Two transactions have a \textit{real-time edge} $\textit{tx}_1 \rt \textit{tx}_2$ 
if there is a real-time ordering between $\textit{tx}_1$ and $\textit{tx}_2$, 
meaning that 
$\textit{tx}_1$ commits before $\textit{tx}_2$'s client issues $\textit{tx}_2$'s first request. 
In an RSG, 
vertices are committed transactions, 
connected by execution and real-time edges. 

There exists a total order if and only if transactions do not circularly affect each other. 
That is, the subgraph that comprises all vertices and only execution edges 
is acyclic, meaning that the following invariant holds: 

\vspace{.5ex}

\invariant{invariant:total}{$\forall \textit{tx}_1,\,\textit{tx}_2\,(\textit{tx}_1 \EX \textit{tx}_2 \implies \neg (\textit{tx}_2 \EX \textit{tx}_1))$}

\vspace{.5ex}

\noindent The (total) execution order respects the real-time order 
if and only if the execution edges (paths) do not \textit{invert} the real-time edges, 
meaning that the following invariant holds:

\vspace{.5ex}

\invariant{invariant:rt}{$\forall \textit{tx}_1,\,\textit{tx}_2\,(\textit{tx}_1 \rt \textit{tx}_2 \implies \neg (\textit{tx}_2 \EX \textit{tx}_1))$}

\vspace{.5ex}

\noindent These invariants correspond to the total order and real-time order 
requirements, respectively. Therefore, a system is strictly serializable if and only if for any execution it allows, 
both invariants hold.  

By enforcing a total order and the real-time order, 
strictly serializable systems provide application programmers with the powerful abstraction of 
programming in a single-threaded, transactionally isolated environment, 
and thus they greatly simplify application development and eliminate consistency anomalies. 
For example, 
if an admin removes Alice from a shared album and then notifies Bob of the change (via a channel external to the system, e.g., a phone call), 
who then uploads a photo he does not want Alice to see, 
then Alice must not see Bob's photo, since $\textit{remove\_Alice} \rt \textit{new\_photo}$. 
Such guarantees cannot be enforced by weaker consistency models, e.g., 
serializability, because they do not enforce the real-time order that is external to the system. 

\begin{figure*}[t]
\centering
\includegraphics[width=.8\linewidth]{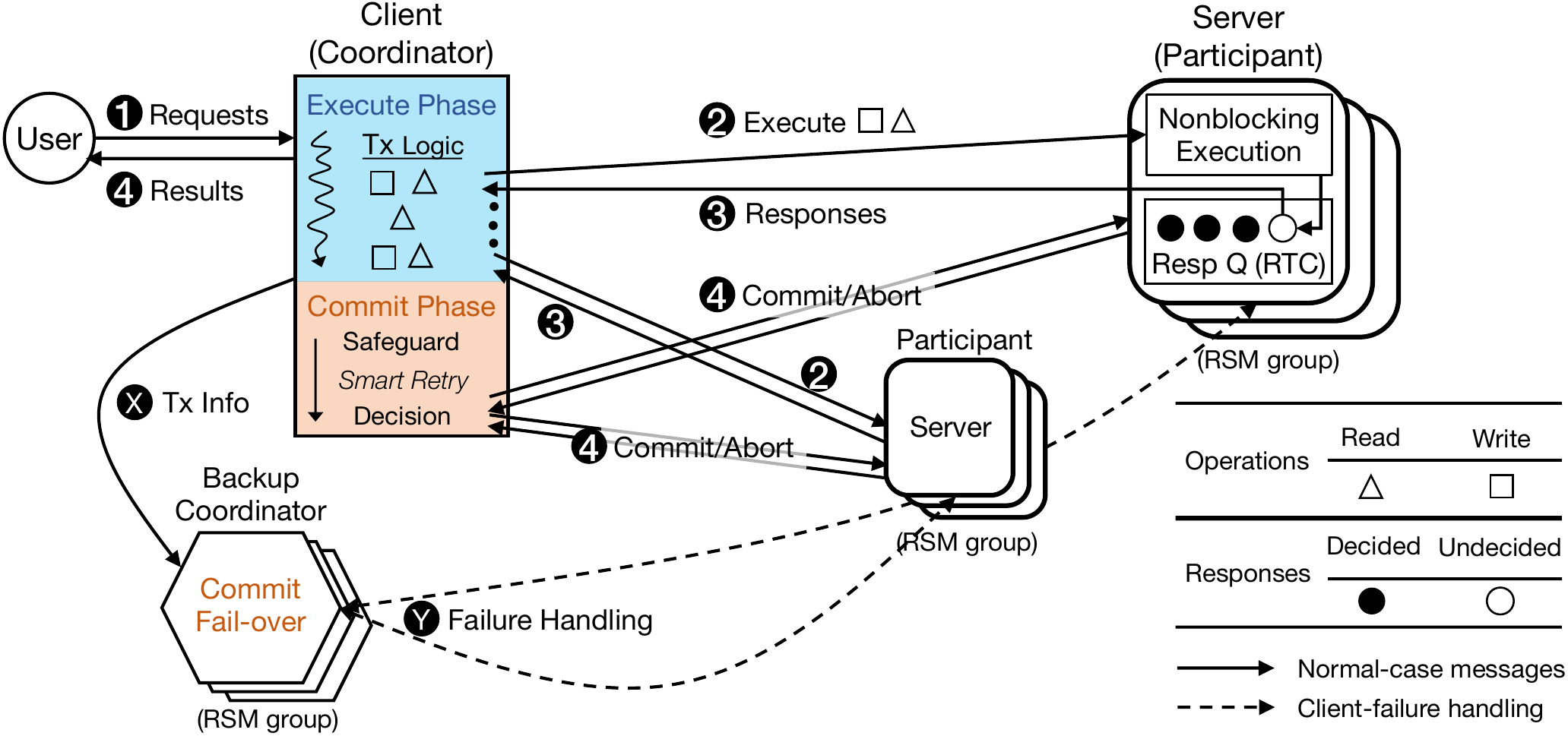}
\caption{An overview of system architecture and transaction execution. 
\sstx{} follows two-phase commit and has   
three design pillars: non-blocking execution, decoupled response management, and 
timestamp-based consistency checking. 
}
\label{fig:arch}
\end{figure*}

\subsection{dOCC, d2PL, \& Transaction Reordering}
\label{subsec:cc}
Only a few techniques provide strict serializability. 
The common ones are dOCC, d2PL, and transaction reordering (TR).
dOCC and d2PL typically require three round trips, one for each phase: 
execute, prepare, and commit.   
In the execute phase, the coordinator reads the data from the servers while writes are buffered 
locally. 
d2PL acquires read locks in this phase while dOCC does not. 
In the prepare phase, 
the coordinator sends prepare messages and the buffered writes to the participant servers. 
d2PL locks all participants while dOCC only locks the written data. 
dOCC must also validate that 
values read in the execute phase have not changed. 
If all requests are successfully prepared, 
i.e., locks are available and/or validation succeeds, 
the coordinator notifies the participants to commit the transaction and apply the writes; 
otherwise, the transaction is aborted and retried.

Transaction reordering typically requires two steps. 
In the first step, the coordinator sends the requests to the servers, 
which make requests wait while recording their arrival order relative to those of concurrent transactions. 
This ordering information usually increases linearly in size with respect to the number of concurrent transactions. 
In the second step, the coordinator collects the ordering information from participants, 
sorts the requests to eliminate interleavings, and servers execute the transactions in the sorted order. 

These techniques are expensive, 
e.g., they require multiple rounds of messages, locking, waiting, and aborts. 
We find that these overheads are wasteful for most of the transactions in 
many datacenter workloads, 
and this observation has inspired our protocol design.

\nps\section{Design Insight \& Overview}
\label{sec:overview}
This section explains natural consistency, which inspires our design, 
and overviews the key design components. 

\subsection{Exploiting Natural Consistency}
\label{subsec:natural}
For many datacenter transactions, simply executing their requests in the order servers receive them, 
as if they were non-transactional read/write operations, would naturally satisfy strict serializability. 
In other words, they arrive at servers in an order that is already strictly serializable. 
We call these transactions \emph{naturally consistent}. 
Key to natural consistency is the arrival order of transaction requests. 

Many requests in datacenter workloads arrive in an order that is total, 
i.e., transactions do not circularly affect each other, 
due to the following reasons. 
First, many requests in real-world workloads are reads~\cite{Corbett:osdi2012, Bronson:atc2013}, 
and reads do not affect other reads. 
For instance, reads that return the same value can be executed in any order, 
and thus servers can safely execute them in their arrival order. 
Second, many transactions are short, 
e.g., they are one-shot~\cite{hstore_one_shot, Mu:osdi2016, Lu:osdi2016, Zhang:sosp2015, kraska2013mdcc, tu2013speedy} or can be made one-shot using stored procedures~\cite{Mu:osdi2014, whitney1997high, li1988multiprocessor, garcia1992main, stonebraker2018end}, 
and thus their requests are less likely to interleave with others' requests. 
Third,  
advances in datacenter networks reduce the variance of message delivery times~\cite{montazeri2018homa, raiciu2019ndp, mittal2015timely}, 
and thus further reduces the likelihood of request interleaving. 

In most cases, the (total) arrival order satisfies the real-time order 
between transactions because a transaction that happens later in real-time, 
i.e., it starts after another transaction has been committed, 
must arrive at servers after the committed transaction has arrived. 

Ideally, the system would treat naturally consistent transactions as non-transactional operations and execute them in the order they arrive without any concurrency control, while still guaranteeing strict serializability.
This insight suggests room for improvement in existing techniques. 
For instance, dOCC still requires validation messages which are unnecessary when  transactions are naturally consistent.
Further, during validation between prepare and commit, dOCC has a \textit{contention window} 
where it can cause other concurrent transactions to abort.
As shown in Figure~\ref{fig:bg-1}, such contention windows lead to \textit{false aborts}, where a transaction is aborted despite being consistent.
Our design 
aims to minimize costs for as many naturally consistent transactions as possible. 

\subsection{Three Pillars of Design}
\label{subsec:pillars}
Our design executes naturally consistent transactions in a manner that closely resembles non-transactional operations. 
This is made possible through three components. 

\paragraph{Non-blocking execution.} Assuming transactions are naturally consistent, 
servers execute requests in the order they arrive. 
Requests are executed ``urgently'' to completion without acquiring locks, 
and their results are immediately made visible to prevent blocking subsequent requests. 
As a result, transactions are executed as cheaply as non-transactional operations, 
without incurring contention windows. 

\paragraph{Decoupled response management.} Because not all transactions are naturally consistent, 
servers must prevent returning inconsistent results to clients 
and ensure there are no cascading aborts. 
This is achieved by decoupling requests' responses 
from their execution, with a response
sent asynchronously only once it is verified consistent. Inconsistent results are discarded, and their requests are re-executed.

\paragraph{Timestamp-based consistency checking.} We must check consistency  as efficiently as possible, 
without interfering with server-side execution. 
We leverage timestamps to capture the arrival order (thus the execution order) of requests 
and design a client-side checker that verifies if requests were executed in a total order, 
without incurring overheads such as messages (as in dOCC and TR) or locks (as in dOCC and d2PL). 

Figure~\ref{fig:arch} shows at a high level how these three pillars support our design,
and depicts the life cycle of transactions: 

\begin{enumerate}[label=(\arabic*),topsep=1pt,itemsep=0ex,partopsep=1ex,parsep=1ex]
    \item[\ding{202}] The user submits application requests to a client, which translates the requests 
    into transactions. 
    \item[\ding{203}] The (client) coordinator sends operations to the participant servers, following the transaction's logic. 
    The servers execute requests in their arrival order.  
    Their responses are inserted into a queue and sent asynchronously. 
    The responses include timestamps that capture requests' execution order. 
    \item[\ding{204}] Responses are sent to the client when it is safe, determined by response timing control (RTC). 
    \item[\ding{205}] The \sg{} checks if transactions were executed in a total order by examining the timestamps in responses. The coordinator sends commit/abort messages to the servers 
    and returns the results of committed transactions to the user in parallel, 
    without waiting for servers' acknowledgments. 
    \mycircled{{\scriptsize X}} and \mycircled{{\scriptsize Y}} explicitly handle client failures by leveraging a server as a backup coordinator.  
\end{enumerate}

\paragraph{Limitations.} 
First, our design leverages natural consistency, 
which is observed in short (e.g., one or few shots) datacenter transactions; 
while our design supports arbitrary-shot transactions,
many-shot long-lasting transactions that are more likely to interleave 
might not benefit from our design.
Second, 
the timestamps associated with each request, including both reads and writes, must be made persistent (e.g., written to disks) and replicated for correctly handling failures, which could lead to replication overhead, which we detail in Section~\ref{subsec:failure}. 

\paragraph{An observation.}
Key to the correctness of our design is leveraging timestamps to verify a total order that respects the real-time order. 
Yet, we identify a correctness pitfall in relying on timestamps to ensure strict serializability.
\nps\section{Timestamp-Inversion Pitfall} 
\label{sec:insight}
We discover that timestamp-based techniques sometimes fail to guard against 
a total order that 
violates the real-time order in subtle cases. 
As a result, executing transactions in such a total order inverts 
the real-time relationship between transactions, which leads to a 
violation of strict serializability. 
We call such violations the \textit{\inversion{}} pitfall. 
Figure~\ref{fig:challenge} shows a minimal construction of 
timestamp inversion 
using three transactions. 
$\textit{tx}_1$ and $\textit{tx}_2$ are single-machine transactions issued by different clients, and $\textit{tx}_2$ starts after $\textit{tx}_1$ finishes, so there exists a real-time
order $\textit{tx}_1 \rt \textit{tx}_2$ 
that strict serializability must enforce. 
$\textit{tx}_3$ is a multi-shard transaction by a third client that interleaves with $\textit{tx}_1$ and $\textit{tx}_2$. 
$\textit{tx}_1$, $\textit{tx}_2$, and $\textit{tx}_3$ have timestamps $10$, $5$, and $7$, respectively.%
\footnote{A timestamp is generated by either a loosely synchronized 
physical clock~\cite{mills1992rfc1305} or a causal counter, e.g., a Lamport clock~\cite{lamport78time}.} 
By following these timestamps, the transactions are executed in a total order denoted as $\textit{tx}_2 \ex \textit{tx}_3 \ex \textit{tx}_1$, 
i.e., $tx_3$ situates itself after $tx_2$ and before $tx_1$, 
which inverts the real-time order $\textit{tx}_1 \rt \textit{tx}_2$ and 
thus violates strict serializability. 
Specifically, the execution of these transactions violates \cref{invariant:rt}, subjecting them to consistency anomalies discussed in \S\ref{subsec:ss}.

The \inversion{} pitfall is subtle because it happens only if a transaction 
interleaves with a set of \emph{non}-conflicting transactions 
that have real-time ordering constraints. 
We find timestamp inversion to be fundamental as it has affected multiple different systems; we discuss two such systems below. In addition, we find that there are several existing systems that do not explicitly define their consistency model, but give a strong indication of providing strict serializability---e.g., they claim  invariants that are equivalent to strict serializability, or are built on or evaluated against strictly serializable protocols. These systems also fall into the pitfall.

\begin{figure}[t]
\centering
\includegraphics[width=0.85\columnwidth]{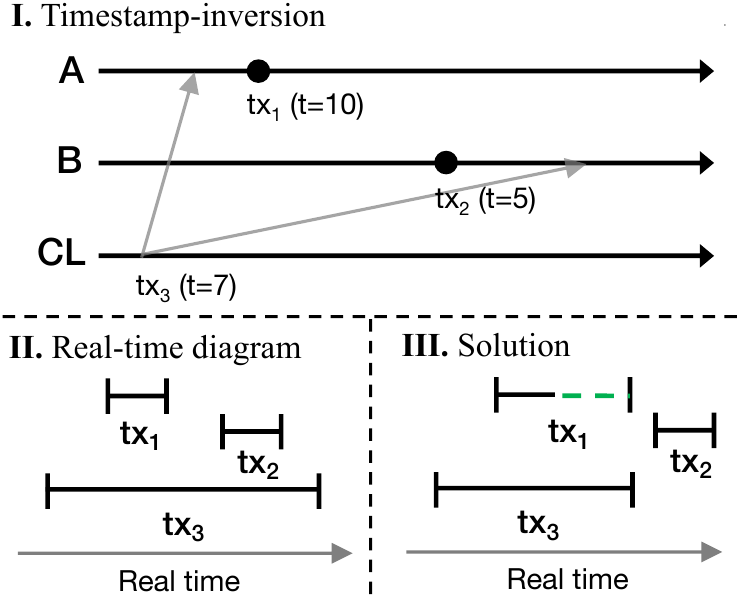}
\caption{A minimal example of timestamp inversion, 
a real-time diagram shows the ordering of transactions,  
and how \sstx tackles the \inversion{} pitfall. 
}
\label{fig:challenge}
\end{figure}

\paragraph{Timestamp inversion affects several prior systems.} 
The minimal example in Figure~\ref{fig:challenge} can be extended to  
variants of timestamp inversion that affect different types of transactions in real system designs, 
suggesting that this pitfall is general and fundamental. 
For instance, we find two systems from recent SOSPs
fall into different variants of the pitfall, 
and thus are not strictly serializable as claimed.
We elaborate below to help future work avoid timestamp inversion, 
and provide the full counterexamples in Appendices~\ref{app:tapir} and~\ref{app:drtm}.

TAPIR~\cite{Zhang:sosp2015, zhang2018building} is an integrated protocol that co-designs concurrency control and replication. 
Its concurrency control is 
a variant of dOCC which 
validates writes using timestamps without acquiring locks, while reads are 
validated in the traditional way. 
Because reads and writes are executed in timestamp order but validated with separate mechanisms, 
TAPIR's read-write transactions may cause an inversion of concurrent writes. 
For instance, if $\textit{tx}_1$, $\textit{tx}_2$, and $\textit{tx}_3$ in Figure~\ref{fig:challenge} are 
read-write transactions, 
then all three transactions would pass TAPIR's validation, which 
results in the inversion of $\textit{tx}_1 \rt \textit{tx}_2$. The effect of this inversion is 
perceivable to the client via future reads. 
This variant of timestamp inversion requires a detailed analysis
of the possible 
executions, 
showing that none of them are admissible by strict serializability (Appendix~\ref{app:tapir}).

DrTM~\cite{Wei:sosp2015, chen2017fast} is a specialized design for 
modern datastores equipped with hardware transactional memory and 
remote direct memory access. 
DrTM uses timestamps to validate read leases which are 
acquired before reading the data, 
a technique equivalent to executing read requests in the timestamp order.
This makes DrTM's read-only transactions subject to inversion, 
e.g., when $\textit{tx}_1$, $\textit{tx}_2$, and $\textit{tx}_3$ in Figure~\ref{fig:challenge} are read-write, read-write, and read-only 
transactions, respectively. 

The main contributions of TAPIR and DrTM still stand, just with weaker consistency than claimed.
Both teams conjecture that they can 
fix the systems 
by using synchronized clocks 
(e.g., TrueTime~\cite{Corbett:osdi2012}) and adapting their designs to use these clocks.
Thus, it is likely that their contributions still stand with strict serializability when synchronized clocks are used.
However, synchronized clocks require specialized infrastructure and are not generally available (\S\ref{sec:related}). 
Therefore, \sstx is designed to avoid \inversion{} without relying on synchronized clocks.

\nps\section{\sysname}
\label{sec:lfc}
This section presents the basic components of \sstx{}, 
explains how \sstx{} avoids the \inversion{} pitfall, 
introduces two timestamp optimization techniques and 
a specialized algorithm for read-only transactions,  
and concludes with discussions of failure handling and correctness. 

\begin{algorithm}[t]
\caption{Client (transaction coordinator) logic} \label{alg:client}
\footnotesize
    \SetKwProg{Fn}{Function}{ :}{end}
    \SetKw{keyin}{in}
    \SetKw{Goto}{go to}
    \SetKw{not}{not}
    \DontPrintSemicolon
    \Fn{\textsc{ExecuteRWTransaction}(tx)} {
        $\textit{results} \gets \{\}$; $\textit{t\_pairs} \gets \{\}$ \tcp{server responses} \label{cli:start}
        \textit{t.clk} $\gets$ \textsc{AsynchronyAwareTS}(\textit{tx}); \textit{t.cid} $\gets$ \textit{clientID}\;
        \label{cl:gen:t} 
        \For {req \keyin{} tx} { \label{cl:send:req1}
            \tcp{send requests shot by shot, \\following tx's logic}
            \textit{res}, \textit{t\_pair} $\gets$ \textsc{NonblockingExecute}(\textit{req}, \textit{t})\; \label{cl:send:req2}
            $\textit{results} \gets \textit{results}\, \cup \,\textit{res}$\;  
            $\textit{t\_pairs} \gets \textit{t\_pairs}\, \cup \,\textit{t\_pair}$
        }
        \tcp{all shots done, tx's logic complete}
        \textit{ok}, $t'$ $\gets$ \textsc{SafeguardCheck}(\textit{t\_pairs})\; \label{cl:sg1}
        \If{\not ok} { \label{cl:sr1}
            \textit{ok} $\gets$ \textsc{SmartRetry}(\textit{tx}, $t'$) \tcp{\S\ref{subsec:sr}} \label{cl:sr2}   
        }
        \If {\textit{ok}} { \label{cl:cmt1}
            \textsc{AsyncCommitOrAbort}(\textit{tx}, ``\textit{committed}'')\;
            \Return \textit{results}\;
        }
        \Else {
            \textsc{AsyncCommitOrAbort}(\textit{tx}, ``\textit{aborted}'')\;
            \Goto~\ref{cli:start} \tcp{abort, and retry from scratch} \label{cl:cmt2}
        }
    }
    \;
    \Fn{\textsc{SafeguardCheck}(t\_pairs)} { \label{cl:sg2}
        $t_w\textit{\_set} \gets \{\}$; $t_r\textit{\_set} \gets \{\}$\;
        \For {\textit{t\_pair} \keyin{} \textit{t\_pairs}} {
            $t_w\textit{\_set} \gets t_w\textit{\_set}\, \cup \,\textit{t\_pair.left}$\;
            $t_r\textit{\_set} \gets t_r\textit{\_set}\, \cup \,\textit{t\_pair.right}$
        }
        $t_w\textit{\_max} \gets \max\{t_w\textit{\_set}\}$; 
        $t_r\textit{\_min} \gets \min\{t_r\textit{\_set}\}$\;
        \If {$t_w\textit{\_max} \le t_r\textit{\_min}$} {
            \tcp{t\_pairs overlap, $\exists$ a snapshot}
            \Return \textit{true}, $t_w\textit{\_max}$\;
        }
        \Else {
            \Return \textit{false}, $t_w\textit{\_max}$\; \label{cl:sg3}
        }
    }   
\end{algorithm}    

\subsection{Protocol Basics}
\label{subsubsec:lowcost}
We build \sstx{} on the three design pillars (\S\ref{subsec:pillars}) to minimize the costs for naturally consistent transactions. 

\paragraph{Pre-timestamping transactions.} 
\sstx{} processes a transaction in two phases: execute and commit. 
Algorithm~\ref{alg:client} shows the client (coordinator)'s logic. 
The coordinator starts a transaction \textit{tx} by pre-assigning it a timestamp \textit{t} 
that consists of two fields: \textit{clk} which is the client's physical time (Section~\ref{subsec:lats} details how it is computed), 
and \textit{cid} which is the client identifier. 
\textit{t} uniquely identifies \textit{tx} (line~\ref{cl:gen:t}). 
When two timestamps have the same \textit{clk}, \sstx{} breaks the tie 
by comparing their \textit{cid}. 
\textit{t} is included in all of \textit{tx}'s requests that are sent to servers 
shot by shot, following \textit{tx}'s application logic (lines~\ref{cl:send:req1} and~\ref{cl:send:req2}). 
These timestamps accompany \textit{tx} throughout its life cycle and will be used to verify if the results are consistent.

\paragraph{Refining timestamps to match execution order.}
Algorithm~\ref{alg:server} details the server-side logic for request execution and commitment. 
Each key stores a list of versions in the order of the server creating them. 
A version has 
three  
fields: \textit{value}, a pair of \textit{timestamps} ($t_w,\,t_r$), and \textit{status}. 
\textit{value} stores the data; 
$t_w$ is the timestamp of the transaction that created the version; 
$t_r$ is the highest timestamp of transactions that read the version; 
and \textit{status} indicates the state of the transaction that created the version: 
either (initially) \textit{undecided}, or \textit{committed}. 
An aborted version is removed from the datastore. 

The server always executes a request against the most recent version \textit{curr\_ver}, 
which is either undecided or committed (line~\ref{svr:ver}). 
Specifically, the server executes a write by creating a new undecided version \textit{new\_ver}, 
which is now the most recent version of the key, ordered after \textit{curr\_ver} (lines~\ref{svr:nv1} and~\ref{svr:nv2}), 
and executes a read by reading the \textit{value} of \textit{curr\_ver} (line~\ref{svr:rvalue}). 
\sstx{}'s basic protocol can work with a single-versioned data store; multi-versioning is required only for \sr{}, a timestamp optimization technique (\S\ref{subsec:sr}). 
The server refines the most recent version's timestamp pair to match the order in which requests are executed. 
Specifically, 
a write request computes \textit{new\_ver}'s $t_w$ as follows: 
its physical time field is no less than that of the write's timestamp $t$ and 
that of \textit{curr\_ver}'s $t_r$, 
and its client identifier is the same as $t$'s (line~\ref{svr:tw}); 
\textit{new\_ver}'s $t_r$ is initialized to $t_w$ (line~\ref{svr:rfw}). 
Similarly, a read request updates \textit{curr\_ver}'s $t_r$ if needed (line~\ref{svr:rfr}). 
Figure~\ref{fig:bg-2} shows examples of how timestamps are refined. 
A version is associated with a $t_w$ and a $t_r$, 
e.g., $A_1$ initially has a timestamp pair ($4,\,8$). 
$\textit{tx}_1$--$\textit{tx}_3$ are single-key read transactions with pre-assigned timestamps $10$, $2$, and $6$, respectively. 
They return the most recent version of $A$, i.e., $A_1$, 
update its $t_r$ if needed, and return $A_1$'s timestamp pair. 
$\textit{tx}_4$ and $\textit{tx}_5$ show how writes manage timestamps. 

These (refined) timestamps match requests' arrival order and thus also match the execution order: 
on each key, a read must have a timestamp greater than that of the write it sees, 
i.e., a read is ordered after the most recent write, 
and a write must have a timestamp greater than that of the most recent read, 
i.e., a write is ordered after the most recent read (and thus all previous writes). 

\begin{algorithm}[t]
\caption{Server execution and commitment} \label{alg:server}
\footnotesize
    \algnewcommand{\And}{\textbf{and}\xspace}
    \algnewcommand{\Or}{\textbf{or}\xspace}
    \SetKwProg{Fn}{Function}{ :}{end}
    \SetKw{keyin}{in}
    \SetKw{Goto}{go to}
    \SetKw{cb}{created by}
    \SetKw{bt}{belongs to}
    \SetKw{is}{is}
    \DontPrintSemicolon
    
    \underline{Multi-versioned data store:}\;
    $\textit{DS}[\textit{key}][\textit{ver}]$    \tcp*{indexed by key, vers sorted by $t_w$\\ver is either committed or undecided}
    \underline{Response queue:}\;
    \textit{resp\_qs}[\textit{key}][\textit{resp\_q}] \tcp*{resp queues for each key}\;    
    \Fn{\textsc{NonblockingExecute}(req, t)} {
        \textit{resp} $\gets$ [ ] \tcp{response message} \label{svr:rsp1}
        $\textit{curr\_ver} \gets \textit{DS}[\textit{req.key}].\textit{most\_recent}$\; \label{svr:ver}
        \If {req \is{} write} {
            $t_w.\textit{clk} \gets \max\{t.\textit{clk},\,\textit{curr\_ver}.t_r.\textit{clk}+1\}$; $t_w.\textit{cid} \gets t.\textit{cid}$\; \label{svr:tw} 
            $t_r \gets t_w$\; \label{svr:rfw}
            \textit{new\_ver} $\gets$ [\textit{req.value}, ($t_w$, $t_r$), \textit{``undecided''}]\; \label{svr:nv1}
            $\textit{DS}[\textit{req.key}] \gets \textit{DS}[\textit{req.key}] + \textit{new\_ver}$\; \label{svr:nv2}
            \textit{resp} $\gets$ [``done'', ($t_w$, $t_r$)]\; \label{svr:rsp2}
        }
        \Else {
            $\textit{curr\_ver}.t_r \gets \max\{t,\,\textit{curr\_ver}.t_r\}$\; \label{svr:rfr}
            \textit{resp} $\gets$ [\textit{curr\_ver}.value, ($\textit{curr\_ver}.t_w$, $\textit{curr\_ver}.t_r$)]\; \label{svr:rvalue}
        }
        \textit{resp\_qs}[\textit{req.key}].enqueue(\textit{resp}, \textit{req}, \textit{t}, ``\textit{undecided}'')\; \label{svr:rsp3}
        \textsc{RespTimingControl}(\textit{resp\_qs}[\textit{req.key}]) \tcp{\S\ref{subsec:avoid-tip}} \label{svr:rsp4}
    }
    \;
    \Fn{\textsc{AsyncCommitOrAbort}(tx, decision)} { \label{svr:commit1}
        \ForEach {ver \cb tx} {
            \If {\textit{decision} $=$ ``\textit{committed}''} {
                \textit{ver.status} $\gets$ \textit{decision}\;
            }
            \Else {
                \textit{DS}.remove(\textit{ver}) \label{svr:commit2}
            }
        }
        \ForEach {resp\_q \keyin{} resp\_qs} { \label{svr:rtc1}
            \ForEach {resp \keyin{} resp\_q} {
                \If {resp.request $\in$ tx} {
                    \textit{resp.q\_status} $\gets$ \textit{decision} \label{svr:rtc2}
                }
            }
            \textsc{RespTimingControl}(\textit{resp\_q}) \tcp{\S\ref{subsec:avoid-tip}} \label{svr:rsp5}
        }
    }
\end{algorithm}

\paragraph{Non-blocking execution and response queues.} 
The server executes requests in a non-blocking manner and decouples their execution from responses. 
Specifically, 
a write creates a version and immediately makes it visible to subsequent transactions; 
a read fetches the value of the most recent version whose \textit{status} could be undecided, 
without waiting for it to commit; 
the server prepares the response (lines~\ref{svr:rsp1},~\ref{svr:rsp2}, and~\ref{svr:rvalue}), inserts it into a \textit{response queue} (lines~\ref{svr:rsp3} and~\ref{svr:rsp4}), which asynchronously sends the responses to clients when it is safe. 
(Section~\ref{subsec:avoid-tip} details response timing control, which determines when sending a response is safe so timestamp inversion and cascading aborts are prevented.)  
Unlike d2PL and dOCC, which lock data 
for at least one round-trip time in the execute and prepare phases 
(i.e., the contention window), 
non-blocking execution ensures that a transaction never exclusively owns 
the data without performing useful work. 
As a result, the server never stalls, and CPUs are fully utilized to execute requests. 
Moreover, non-blocking execution eliminates the contention window 
and thus reduces false aborts. 

\paragraph{Client-side \sg{}.} 
A server response includes the timestamp pair ($t_w,\,t_r$) of the 
most recent version, e.g., \textit{new\_ver} for a write and \textit{curr\_ver} for a read. 
The returned ($t_w,\,t_r$) represents the time range in which the request is valid. 
That is, a read must take effect after $t_w$, which is the time when the most recent write on the same key took effect, and no later writes can take effect between $t_w$ and $t_r$ on the same key. 
A write must have $t_w = t_r$, meaning that it takes effect exactly at $t_w$. 
When a transaction has completed its logic (i.e., all shots are executed) and the client has 
received responses to all its requests, the \sg{} looks for 
a consistent snapshot that intersects all ($t_w, \,t_r$) pairs in server responses 
by checking if the ($t_w,\,t_r$) pairs overlap (lines~\ref{cl:sg1},~\ref{cl:sg2}--\ref{cl:sg3}). 
This intersecting snapshot identifies the transaction's synchronization point, 
i.e., all requests are valid at the intersecting timestamp. 

Figure~\ref{fig:bg-3} shows an example where \sstx executes 
the same transactions in Figure~\ref{fig:bg-1}. 
The default versions $A_0$ and $B_0$ both have a timestamp pair ($0,\,0$). 
$tx_1$ and $tx_2$ are pre-assigned $4$ and $8$, respectively, 
and their requests arrive in the same order as they were in Figure~\ref{fig:bg-1}. 
The \sg{} enables \sstx{} to commit both transactions, 
i.e., $tx_1$'s responses intersect at $4$ while $tx_2$'s responses intersect 
at $8$, 
without unnecessary overhead such as dOCC's validation cost and false aborts. 

When the client has decided to commit or abort the transaction, the protocol enters the commit phase by 
sending the commit/abort messages to the servers. 
If the transaction is committed, the server updates the \textit{status} of the created versions 
from undecided to committed; 
otherwise, the versions are deleted (lines~\ref{svr:commit1}--\ref{svr:commit2}). 
The client retries the aborted transaction. 
The client sends the results of the committed transaction to the user 
in parallel with the commit messages, 
i.e., asynchronous commit, 
without waiting for servers' acknowledgments (lines~\ref{cl:cmt1}--\ref{cl:cmt2}). 

\paragraph{Supporting complex transaction logic.} 
\sstx{}
supports 
transactions accessing a key multiple times, e.g., 
read-modify-writes and repeated reads/writes, 
by treating its requests to the same key as a single logical request. 
For instance, 
if a read-modify-write has its read and write requests executed consecutively 
(i.e., they are not intersected by other writes), 
then only the write response is checked by the \sg, 
treating read-modify-write as one logical request; 
otherwise, it is aborted if there are intersecting writes, 
e.g., when the most recent version has a $t_w$ greater than that returned by the read of 
this read-modify-write.  
The responses of these requests are grouped together in the response queue, e.g., 
the write response of a read-modify-write is inserted right after the read response of the same read-modify-write. 
We explain the details of handling complex logic in Appendix~\ref{app:detail}. 

\sstx{} achieves minimal costs by urgently executing transactions 
in a non-blocking manner and by ensuring a total order with the light-weight timestamp-based \sg{}. 
Yet, in order to provide strict serializability, \sstx{} must enforce the real-time order between transactions by handling the \inversion{} pitfall, 
as we discuss next. 

\subsection{Response Timing Control}
\label{subsec:avoid-tip}
\sstx{} avoids the \inversion{} pitfall by disentangling the subtle interleaving between a set of non-conflicting transactions that have real-time order dependencies (e.g., Figure~\ref{fig:challenge}), without relying on synchronized clocks. 
Specifically, \sstx{} introduces \textit{response timing control} (RTC), 
which controls the sending time of responses. 
It is safe to send the response of a request $\textit{req}_1$ when the following dependencies are satisfied: 

\begin{enumerate}[label=(\arabic*),topsep=1pt,itemsep=0ex,partopsep=1ex,parsep=1ex]
    \item[\setword{D\textsubscript{1}}{dep1}] If $\textit{req}_1$ reads a version created by $\textit{req}_0$ of another transaction, then $\textit{req}_1$'s response is not returned until $\textit{req}_0$ is committed or it is discarded  if $\textit{req}_0$ is aborted (then $\textit{req}_1$ will be re-executed). 

 \item[\setword{D\textsubscript{2}}{dep2}] If $\textit{req}_1$ is a write and there are reads that read the version which immediately precedes the one created by $\textit{req}_1$, then $\textit{req}_1$'s response is not returned until the reads are committed/aborted.  
	\item[\setword{D\textsubscript{3}}{dep3}] If $\textit{req}_1$  
 creates a version  immediately after the version created by $\textit{req}_0$ of another transaction, then $\textit{req}_1$'s response is not returned until $\textit{req}_0$ is committed/aborted.  
\end{enumerate}
 
By enforcing these dependencies, 
\sstx{} controls the sending of responses 
so that the transactions which form the subtle interleaving 
are forced to take effect in their real-time order. 
For instance, in Figure~\ref{fig:challenge}, 
server $A$ cannot send the response of $\textit{tx}_1$ 
until $\textit{tx}_3$ has been committed 
(assuming at least one of them writes to $A$). 
As a result, any transaction $\textit{tx}_2$ that begins after $\textit{tx}_1$ receives its response, 
i.e., 
$\textit{tx}_1 \rt \textit{tx}_2$, 
must be executed after $\textit{tx}_1$, 
and thus after $\textit{tx}_3$ as well. In detail,
$\textit{tx}_2$'s execution on each server is after it begins,
which is after $\textit{tx}_1$ ends, 
which is after $\textit{tx}_1$'s response is sent, 
which is after $\textit{tx}_3$ commits,
which is after $\textit{tx}_3$ executes on each server.
This results in a total order 
$\textit{tx}_3 \ex \textit{tx}_1 \ex \textit{tx}_2$, 
which respects the real-time order of
\cref{invariant:rt}, as shown in Part~\ROM{3} of Figure~\ref{fig:challenge}. 

\sstx{} implements RTC by managing response queues 
independently from request execution. 
\sstx{} maintains one queue per key. 
A queue item consists of four fields: 
\textit{response} that stores the response message of a request, 
the \textit{request} itself, 
\textit{ts} which is the pre-assigned timestamp of the request, 
and \textit{q\_status} that indicates the state of the request, 
which is initially \textit{undecided}, and updated to either \textit{committed} or \textit{aborted} 
when the server receives the commit/abort message for this request (lines~\ref{svr:rtc1}--\ref{svr:rtc2}, Algorithm~\ref{alg:server}).

\begin{algorithm}[t]
\caption{Response timing control} \label{alg:rtc}
\footnotesize
    \algnewcommand{\And}{\textbf{and}\xspace}
    \algnewcommand{\Or}{\textbf{or}\xspace}
    \SetKwProg{Fn}{Function}{ :}{end}
    \SetKwProg{Infloop}{repeated loop}{}{end}
    \SetKw{keyin}{in}
    \SetKw{Goto}{go to}
    \SetKw{cb}{created by}
    \SetKw{bt}{belongs to}
    \SetKw{is}{is}
    \SetKw{isnot}{is not}
    \SetKw{syscall}{sys\_call}
    \SetKw{bk}{break repeated loop}
    \DontPrintSemicolon
    
    \Fn{\textsc{RespTimingControl}(resp\_q)} {        
        \textit{head} $\gets$ \textit{resp\_q}.head() \tcp{the oldest response}\label{rtc:1}
        \While {\textit{head.q\_status} $\neq$ ``\textit{undecided}''} {\label{rtc:itr1}
            \tcp{find the first response we can send}
            \textit{resp\_q}.dequeue()\; \label{rtc:2}
            \textit{new\_head} $\gets$ \textit{resp\_q}.head()\;
            \textit{new\_req} $\gets$ \textit{new\_head.request}; \textit{t} $\gets$ \textit{new\_head.ts}\; 
            \While {head.q\_status $=$ ``aborted''\; \label{rtc:fix1} \textbf{and} head.request \is{} write \textbf{and} new\_req \is{} read} {
                \tcp{handle reads seeing aborted writes}
                \textit{resp\_q}.dequeue() \tcp{discard read response} 
                \tcp{re-execute the read locally}
                \textsc{NonblockingExecute}(\textit{new\_req}, \textit{t})\; \label{rtc:fix2}
                \textit{new\_head} $\gets$ \textit{resp\_q}.head()\;
                \textit{new\_req} $\gets$ \textit{new\_head.request}; \textit{t} $\gets$ \textit{new\_head.ts}\;
            }
            \textit{head} $\gets$ \textit{resp\_q}.head() \label{rtc:itr2}
        }
        \textit{curr\_item} $\gets$ \textit{head}\; \label{rtc:3}
        \Infloop{}{\label{rtc:5}
            \tcp{send dependency-satisfied responses}
            \textit{resp} $\gets$ \textit{curr\_item.response}\; \label{rtc:4}
            \If {resp.is\_sent $\neq$ true} {\label{rtc:send1}
                \syscall{}.send(\textit{resp}) \tcp{send to client}
                resp.is\_sent $\gets$ true \label{rtc:send2}    
            }
            \tcp{send consecutive read responses}
            \textit{next\_item} $\gets$ \textit{curr\_item}.next()\; \label{rtc:read1}
            \If {curr\_item.request \isnot{} read \\ \textbf{or} next\_item.request \isnot{} read} {
                \bk{} \label{rtc:read2}
            }
            \textit{curr\_item} $\gets$ \textit{next\_item}
        }   
    }

\end{algorithm}

\paragraph{Managing response queues.} 
Algorithm~\ref{alg:rtc} details how \sstx{} manages the response queue of each key. 
This logic is invoked every time the server finishes executing a request (line~\ref{svr:rsp4}) and receives a commit/response message (line~\ref{svr:rsp5}). 
\sstx{} 
iterates over the queue items from the head (i.e., the oldest response) 
until it finds the first response whose \textit{q\_status} is undecided, 
which means 
all earlier requests on the same key have been committed or aborted, 
i.e., this response has satisfied the three dependencies 
(lines~\ref{rtc:1}--\ref{rtc:2} and~\ref{rtc:itr2}). 
The server sends this response message to the client if it has not done so (lines~\ref{rtc:3},~\ref{rtc:4}--\ref{rtc:send2}). 
If this is a read response, then the server sends all consecutive read responses that follow it (lines~\ref{rtc:5} and~\ref{rtc:read1}--\ref{rtc:read2}), 
because all these read responses satisfy 
the three dependencies. 
In other words, reads returning the same value do not have dependencies between them. 
RTC is \textit{effectively} similar to locking the response queues, 
e.g., the queue is ``locked'' when a response is sent and other responses 
must wait, and is ``unlocked'' when the commit/abort message for the request to which the sent response belongs is received. 
However, RTC differs from lock-based mechanisms in that  
it is decoupled from execution and does not introduce contention windows, 
i.e., data objects are not locked.  

\paragraph{Fixing reads locally.} 
When the server receives an abort message for a write request, it must invalidate the responses of any reads that have fetched the value of the aborted write. This is necessary to avoid returning invalid results to the client and to prevent cascading aborts. 
Specifically, 
the server removes the response of such a read from the response queue and 
re-executes the read request, 
e.g., it fetches the current most recent version,  
prepares a new response, and inserts the new response to the tail of the queue (lines~\ref{rtc:fix1}--\ref{rtc:fix2}). 

\paragraph{Avoiding indefinite waits.} 
To avoid responses from circularly waiting on dependencies across different keys, 
\sstx{} early aborts a request (thereby aborting the transaction to which it belongs) 
if its pre-assigned timestamp is not the highest the server has seen \textit{and} if its response cannot be sent immediately, 
i.e., it is not the head of the queue. 
Specifically, 
a write (read) is aborted if there is an undecided request (write request) with a higher timestamp. 
Then, the server sends a special response to the client without 
executing the request. 
The special response includes a field \textit{early\_abort} which allows 
the client to bypass the \sg{} and abort the transaction. 
We omit the details from the pseudocode for clarity. 

RTC is a general solution to timestamp inversion, 
without the need for synchronized clocks. 
It does not incur more aborts even when responses are not sent immediately, 
because response management is decoupled from request execution. 
That is, whether a transaction is committed or aborted is solely based on 
timestamps, and RTC does not affect either pre-assignment or refinement of timestamps. 
Yet, \sstx{}'s performance also depends on how well timestamps capture 
the arrival order of (naturally consistent) transactions. 
That is, timestamps that do not match transactions' 
arrival order could cause transactions to falsely abort even if they are naturally consistent. 
Next, we will discuss optimization techniques that 
enable timestamps to better match the arrival order. 

\subsection{Asynchrony-Aware Timestamps}
\label{subsec:lats}
\sstx{} proposes two optimizations: a proactive approach that controls how timestamps 
are generated before transactions start, 
and a reactive approach that updates timestamps to match the naturally consistent arrival order after requests are executed. 
This subsection discusses the proactive approach. 

The client pre-assigns the same timestamp to all requests of a transaction; 
however, these requests may arrive at their participant servers at different physical times, which could result in a mismatch between timestamp and arrival order, 
as shown in Figure~\ref{fig:opt-1}. 
Transactions $\textit{tx}_1$ and $\textit{tx}_2$ start around the same time and thus 
are assigned close timestamps, e.g., 
$t_1=1004$ and $t_2=1005$, respectively (client IDs are omitted). 
Because the latency between $B$ and $\textit{CL}_1$ is greater than that between $B$ and $\textit{CL}_2$, 
$\textit{tx}_1$ may arrive at $B$ later than $\textit{tx}_2$, but $\textit{tx}_1$ has 
a smaller timestamp. 
As a result, 
the \sg may falsely reject $\textit{tx}_1$, 
e.g., server $B$ responds with a refined timestamp pair ($1006,\,1006$) 
which does not overlap with ($1004,\,1004$), the timestamp pair returned by server $A$. 
However, aborting $\textit{tx}_1$ is unnecessary because 
$\textit{tx}_1$ and $\textit{tx}_2$ are naturally consistent. 

To tackle this challenge, \sstx{} generates timestamps while accounting for the time difference, $t_\Delta$, between when a request is sent by the client and when the server starts executing the request. 
Specifically, 
the client records the physical time $t_c$ before sending the request to the server; 
the server records the physical time $t_s$ before executing the request and piggybacks $t_s$ onto the response sent back to the client; and 
the client calculates $t_\Delta$ by finding the difference between $t_c$ and $t_s$, 
i.e., $t_\Delta=t_s-t_c$.
By measuring the end-to-end time difference, 
$t_\Delta$ effectively masks the impact of 
queuing delays and clock skew. 
The client maintains a $t_\Delta$ for each server it has contacted. 
An asynchrony-aware timestamp is generated by adding the client's current physical time and the greatest $t_\Delta$ among the servers this transaction will access. 
For instance, given the values of $t_\Delta$ shown in 
Figure~\ref{fig:opt-1}, 
$\textit{CL}_1$ assigns $\textit{tx}_1$ timestamp $1014$ 
(i.e., $1004+10$) 
and $\textit{CL}_2$ assigns $\textit{tx}_2$ $1010$ (i.e., $1005+5$), and both transactions may successfully pass their \sg{} check,
availing their natural consistency. 

\begin{figure*}[t]
\centering
\begin{subfigure}[b]{0.3\linewidth}
  \centering
  \includegraphics[width=1.0\linewidth]{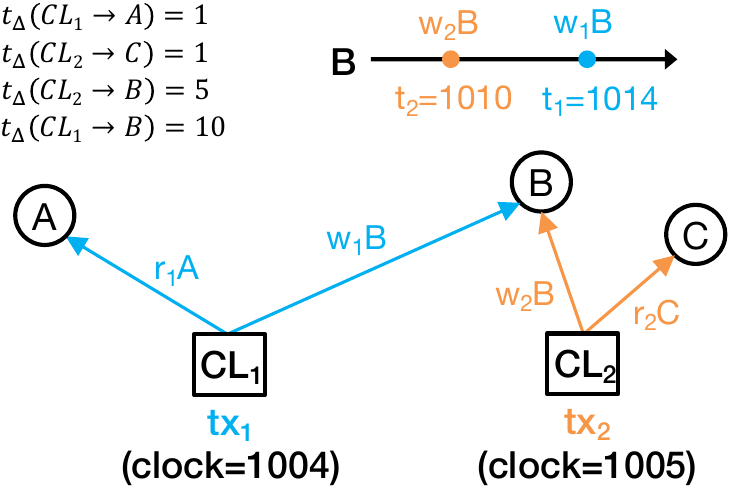}
  \caption{Asynchrony-aware timestamps}
  \label{fig:opt-1}
\end{subfigure}
~~~~~~~~
\begin{subfigure}[b]{0.3\linewidth}
  \centering
  \includegraphics[width=1.0\linewidth,page=2]{figs/opt.pdf}
  \caption{\Sg{} false rejects}
  \label{fig:opt-2}
\end{subfigure}
~~~~~~~~
\begin{subfigure}[b]{0.3\linewidth}
  \centering
  \includegraphics[width=1.0\linewidth,page=3]{figs/opt.pdf}
  \caption{Smart retry reduces false rejects}
  \label{fig:opt-3}
\end{subfigure}
\caption{Optimizations that match the timestamps with 
transactions' arrival order. 
Asynchrony-aware timestamps proactively controls the pre-assigned timestamps before execution. 
\Sr{} reactively fixes the \sg{}'s false rejects after execution thus avoids aborting and re-executing transactions.}
\label{fig:opt}
\end{figure*}

\subsection{Smart Retry}
\label{subsec:sr} 

\begin{algorithm}[t]
\caption{\Sr{}} \label{alg:sr}
\footnotesize
    \SetKwProg{Fn}{Function}{ :}{end}
    \SetKw{cb}{created by}
    \SetKw{rb}{read by}
    \SetKw{ab}{accessed by}
    \DontPrintSemicolon  

    \Fn{\textsc{SmartRetry}(tx, $t'$)} { 
        \ForEach {ver \ab{} tx} {
            \tcp{next version of the same key}
            \textit{next\_ver} $\gets$ \textit{ver}.next()\; \label{sr:f1}
            \If {next\_ver.$t_w$ $\le$ $t'$} { \label{sr:f2}
                \Return \textit{false}\; \label{sr:f3}
            }
            \If {ver \cb{} tx \textbf{and} \textit{ver}.$t_w$ $\neq$ \textit{ver}.$t_r$} { \label{sr:w1}
                \Return \textit{false}\; \label{sr:w2} 
            }
            \If {ver \cb{} tx} { \label{sr:update1}
                \textit{ver}.$t_w$ $\gets$ $t'$; \textit{ver}.$t_r$ $\gets$ $t'$\; 
            }
            \Else {
                \textit{ver}.$t_r$ $\gets$ $max\{\textit{ver}.t_r,\,t'\}$\; \label{sr:update2}
            }
        }
        \Return \textit{true}\;
    }
\end{algorithm}

\sstx{} proposes a reactive approach to minimizing the performance impact of 
the \sg{}'s false rejects, 
which happen when timestamps fail to identify the naturally consistent arrival order, 
as shown in Figure~\ref{fig:opt-2}. 
Initially, version $A_0$ has a timestamp pair ($0,\,0$), and $B_0$ has ($0,\,5$). 
The same transactions $\textit{tx}_1$ and $\textit{tx}_2$ as those in 
Figure~\ref{fig:bg-3} access both keys. 
Following \sstx{}'s protocol, $\textit{tx}_1$'s responses contain the timestamp pairs ($0,\,4$) and ($6,\,6$) from $A$ and $B$, respectively, 
which will be rejected by the \sg{} because they do not overlap. 
However, aborting $\textit{tx}_1$ is unnecessary because $\textit{tx}_1$ and $\textit{tx}_2$ are naturally consistent. 

Instead, 
\sstx{} tries to ``reposition'' a rejected transaction with respect to the transactions before and after it to construct a total order, instead of 
aborting and re-executing the rejected transaction from scratch, which would 
waste all the work the server has done for executing it. 
Specifically, \sstx{} chooses a timestamp that is nearest ``in the future'' and hopes the rejected transaction can be re-committed at that time. 
This is possible if the chosen time has not been taken by other transactions.  

Algorithm~\ref{alg:sr} shows the pseudocode for \sr{}. 
When the transaction fails the \sg{} check, 
\sstx suggests a new timestamp $t'$, 
which is the maximum $t_w$ in the server responses. 
The client then sends \sr{} messages that include $t'$ to the participant servers, which then attempt to reposition the transaction's requests at $t'$. 
The server can reposition a request if there has not been a newer version that was created before $t'$ (lines~\ref{sr:f1}--\ref{sr:f3}) and, 
if the request is a write, 
the version it created has not been read by any transactions (lines~\ref{sr:w1} and~\ref{sr:w2}). 
The server updates the timestamps of relevant versions if \sr{} succeeds, 
e.g., the created version has a new timestamp pair ($t',\,t'$), 
and $t_r$ of the read version is updated to $t'$ if $t'$ is greater (lines~\ref{sr:update1}--\ref{sr:update2}). 
(Our implementation does not smart-retry the request that returned the maximum $t_w$, i.e., $t_w=t'$, because its \sr{} always succeeds.)
The client commits the \sg{}-rejected transaction if all its \sr{} requests succeed, 
and aborts and retries it from scratch 
otherwise (lines~\ref{cl:sr1} and~\ref{cl:sr2}, Algorithm~\ref{alg:client}). 

Not only does \sr{} avoid 
false aborts, 
it also unleashes a higher degree of concurrency,  
as shown in Figure~\ref{fig:opt-3}.  
The servers have executed a newer transaction $\textit{tx}_2$ when 
$\textit{tx}_1$'s smart retry (SR) messages arrive, and 
both transactions can be committed even if the messages interleave, 
e.g., $\textit{tx}_1$'s \sr{} succeeds and $\textit{tx}_2$ passes its \sg{} 
check, 
because $\textit{tx}_2$'s pre-assigned timestamps have left enough room 
for repositioning $\textit{tx}_1$'s requests. 
In contrast,  
validation-based techniques would unnecessarily abort $\textit{tx}_1$ 
(considering SR as dOCC's validation messages) due to the presence of the 
conflicting transaction $\textit{tx}_2$. 

\paragraph{Garbage collection.} 
Old versions are temporarily stored 
and garbage collected as soon as 
they are no longer needed by undecided transactions for \sr{}. 
Only the most recent versions are used to serve new transactions. 

\subsection{Read-Only Transactions}
\label{subsec:ro}
\sstx designs a specialized read-only transaction protocol for 
read-dominated workloads~\cite{Bronson:atc2013, Corbett:osdi2012, Lu:sosp2015, Lu:osdi2016, konwar2021snow}.  
Similar to existing works, \sstx{} 
optimizes 
read-only transactions by eliminating their commit phase because they do not modify the system state and have nothing to commit.  
By eliminating commit messages,  
read-only transactions achieve \textit{optimal performance} in the best case, 
i.e., one round of non-blocking messages with constant metadata~\cite{Lu:osdi2016, lu2020performance, lu:osdi20tr}. 

Eliminating commit messages 
brings a new challenge to 
response timing control: 
write responses can no longer track their dependencies on preceding read-only transactions, 
as they do not 
know if and when those reads are committed/aborted. 
To tackle this challenge,  
\sstx aborts a read-only transaction if it could possibly cause the subtle interleaving that leads to timestamp inversion. 
In other words, \sstx{} commits a read-only transaction if its requests arrive in 
a naturally consistent order and no intervening writes have been executed since the last time the client accessed these servers. 

Specifically, each client  
tracks $t_{ro}$ which is the $t_w$ of 
the version created by the most recent write 
on a server, 
and the client maintains a map of $t_{ro}$ for each server  
this client has contacted. 
A read-only transaction is identified by a Boolean field \textit{IS\_READ\_ONLY}. 
The client sends each of its requests to the participant server together 
with the pre-assigned timestamp (as in the basic protocol) and the $t_{ro}$ of the server. 
To execute a read request, the server checks the version at $t_{ro}$. 
If the version is still the most recent, the server continues to execute the read following the basic protocol,  
i.e., it fetches the most recent version, refines its $t_r$ if needed, 
and returns its timestamp pair; 
otherwise, the server sends a special response that contains a field \textit{ro\_abort} immediately without executing the request. 
If any of the responses contain \textit{ro\_abort}, the client 
aborts this read-only transaction; 
otherwise, the client continues with the \sg{} check and, if needed, \sr{}, 
after which the client does not send any commit/abort messages. 

This protocol pays more aborts in the worst case in exchange for reduced message overhead in the normal case, 
a trade-off that is worthwhile for read-dominated workloads 
where writes are few (so aborts are rare), and read-only transactions are many, making 
the savings in message cost significant. 
This protocol also expedites the sending of responses for read-write transactions 
because read-only transactions do not insert responses into the response queue, 
i.e., a write response depends only on the reads of preceding read-write transactions in 
Dependency~\ref{dep2}, not those of read-only transactions. 

\subsection{Failure Handling}
\label{subsec:failure}
\paragraph{Tolerating server failures.} 
\sstx{} assumes servers never fail as 
their state is typically 
made persistent on disks and replicated via state machine replication 
such as Paxos~\cite{leslie1998part}. 
All state changes incurred by a transaction in the execute phase 
(e.g., $t_w$ and $t_r$ of each request) 
must be written to the disk and replicated for correctness. 
For instance, 
after a request is executed, the server inserts its response into the response queue 
and, in parallel, writes the state changes to the disk and replicates the request to other replicas. 
Its response is sent back to the client when it is allowed by response timing control and 
when its replication is finished. 
Commit/abort and \sr{} messages are also made persistent and replicated. 
This simple scheme ensures correctness but incurs high overhead. 
We plan to investigate possible optimizations in future work, 
e.g., \sstx{} could defer disk writes and replication to the last shot of a transaction where 
all state changes are made persistent and replicated once and for all, 
without having to replicate each request separately. 
Server replication inevitably increases latency but does not introduce more aborts, 
because whether a transaction is committed or aborted is solely based on its timestamps, 
which are decided during request execution and before replication starts. 

\paragraph{Tolerating client failures.} 
\sstx{} must handle client failures explicitly 
because clients are not replicated in most systems 
and \sstx{} co-locates coordinators with clients. 
\sstx{} adopts an approach similar to that in Sinfonia~\cite{Aguilera:sosp2007} and RIFL~\cite{Lee:sosp2015}.  
We briefly explain it as follows and present its details in Appendix~\ref{app:fh}.  
For a transaction \textit{tx}, one of the storage servers \textit{tx} accesses is selected 
as the backup coordinator, 
and the other servers are cohorts. 
In the last shot of the transaction logic, which is identified by a field \textit{IS\_LAST\_SHOT} in the requests, the client notifies the backup coordinator of the identities of the complete set of cohorts. 
Cohorts always know which server is the backup coordinator. 
When the client crashes, e.g., is unresponsive for a certain amount of time, 
the backup coordinator reconstructs the final state of $\textit{tx}$ 
by querying the cohorts for how they executed $\textit{tx}$, 
and commits/aborts \textit{tx} following the same \sg and \sr{} logic.
Because this computation is deterministic, the backup coordinator always makes the same commit/abort 
decision as the client would if the client did not fail. 
To tolerate one client failure, \sstx{} needs one backup coordinator which is selected among storage servers replicated in a usual way. 

\subsection{Correctness}
\label{subsec:correctness}
This section provides proof intuition for why \sstx is safe and live. 
At a high level, \sstx{} guarantees a total order, the real-time order, and liveness, 
with the mechanisms (M\textsubscript{1}) the \sg{}, (M\textsubscript{2}) non-blocking execution with response timing control, 
and (M\textsubscript{3}) early aborts, respectively. 
We provide a formal proof of correctness in Appendix~\ref{app:proof}.

\paragraph{\sstx{} is safe.} 
We prove that \sstx{} guarantees strict serializability by 
demonstrating that 
both \cref{invariant:total,invariant:rt} are upheld. 
These two invariants correspond to the total order and real-time order requirements, respectively. 

Intuitively, \sstx{} commits all requests of a transaction at the same synchronization point, 
which is the intersection of all ($t_w,\,t_r$) pairs in responses, 
and the synchronization points of all committed transactions construct a total order. 
Specifically, we prove that the \sg{} enforces \cref{invariant:total}, by contradiction. 
Assume both $tx_1$ and $tx_n$ are committed, and $\textit{tx}_1 \EX \textit{tx}_n \EX \textit{tx}_1$. 
Without loss of generality, there must exist
a chain of transactions such that $\textit{tx}_1 \ex \textit{tx}_2 \ex \ldots \ex \textit{tx}_n \ex \textit{tx}_1$. 
Then, consider requests $\textit{req}$ and $\textit{req}'$ 
in each pair of adjacent transactions, 
we must have 
$\textit{req}_1' \ex \textit{req}_2$, $\textit{req}_2' \ex \textit{req}_3$, $\ldots\,$, $\textit{req}_{n-1}' \ex \textit{req}_n$, $\textit{req}_n' \ex \textit{req}_1$. 
Considering their returned timestamps, we can derive the following: 
\begin{enumerate}[label=(\arabic*),topsep=1pt,itemsep=0ex,partopsep=1ex,parsep=1ex]
    \item[\circled{1}] $t_{r1}' \le t_{w2}$, $t_{r2}' \le t_{w3}$, $\ldots\,$, $t_{rn}' \le t_{w1}$, by \sstx{}'s protocol. 
    \item[\circled{2}] $t_{w1} \le t_{r1}'$, $t_{w2} \le t_{r2}'$, $\ldots\,$, $t_{wn} \le t_{rn}'$, because all transactions are committed and by the \sg{} logic.
    \item[\circled{3}] $t_{w1} \le t_{r1}' \le t_{w2} \le t_{r2}' \le \ldots \le t_{wn} \le t_{rn}' \le t_{w1}$, by~\circled{1} and~\circled{2}. 
    \item[\circled{4}] $t_{r1}' = t_{w2} = t_{r2}' = t_{w3} = \ldots = t_{wn} = t_{rn}' = t_{w1}$, by~\circled{3}. 
    \item[\circled{5}] $\textit{req}_i'$ is a write and $\textit{req}_i$ is a read, $i \in [1,\,n]$,  
    by~\circled{4}, \sstx{}'s protocol, and $\textit{tx}_1 \ex \textit{tx}_2 \ex \ldots \ex \textit{tx}_n \ex \textit{tx}_1$. 
    \item[\circled{6}] $t_{w2} = t_{w1}'$ and $t_{w1} = t_{wn}'$, by \circled{5} and \sstx{}'s protocol. 
    \item[\circled{7}] $t_{w1}' = t_{wn}'$, by~\circled{4} and~\circled{6}, which contradicts that writes from different transactions must have distinct $t_w$ because timestamps are unique. Therefore, \cref{invariant:total} holds.  
\end{enumerate}

We prove that \sstx enforces \cref{invariant:rt} by assuming $\textit{tx}_1 \rt \textit{tx}_2$ and showing that it must be that $\textit{tx}_1 \EX \textit{tx}_2$.
There are two cases to consider.
In case 1, $\textit{tx}_1$ and $\textit{tx}_2$ 
access some common data items. 
Then, we must have $\textit{tx}_1 \EX \textit{tx}_2$, 
because \sstx{} executes requests in their arrival order. 
Then, it must be true that $\neg (\textit{tx}_2 \EX \textit{tx}_1)$, by \cref{invariant:total}. 
In case 2, $\textit{tx}_1$ and $\textit{tx}_2$ access disjoint data sets, and we prove the claim by contradiction. 
Assume that $\textit{tx}_2 \EX \textit{tx}_1$; 
then, there must exist $\textit{req}_2$ and $\textit{req}_1$ in $\textit{tx}_2$ and $\textit{tx}_1$, respectively, such that $\textit{req}_2 \EX \textit{req}_1$. 
$\textit{req}_1$'s response is not returned until $\textit{req}_2$ is committed or aborted, by applying response timing control (\S\ref{subsec:avoid-tip}). 
Then, $\textit{req}_2$ is issued before $\textit{req}_1$'s client receives $\textit{req}_1$'s response because a request, e.g., $\textit{req}_2$, can be committed or aborted only after it is issued.
Thus, we can derive $\neg (\textit{tx}_1 \rt \textit{tx}_2)$ because $\textit{tx}_2$ has at least one request, e.g., $\textit{req}_2$, which starts before $\textit{tx}_1$ receives all its responses. 
This means $\textit{tx}_2$ starts before $\textit{tx}_1$ is committed, 
which contradicts our assumption $\textit{tx}_1 \rt \textit{tx}_2$. 
Therefore, \cref{invariant:rt} must hold.  

\begin{figure*}[t]
\footnotesize
\centering
\renewcommand{\arraystretch}{1.5}
\begin{tabular}{@{}l cc  ccc cc@{}}
\Xhline{1.5pt}
{\bf Workload} & {\bf Write fraction} & {\bf Assoc-to-obj} & {\bf \# keys/RO} & {\bf \# keys/RW} & {\bf Value size} & {\bf \# cols/key} & {\bf Zipfian} \\
\hline
Google-F1  & $0.3\%$ [$0.3\%$--$30\%$] & --- & $1$--$10$ & $1$--$10$ & $1.6\,\textrm{KB} \pm 119\,\textrm{B}$ & $10$ & $0.8$\\
\hline
Facebook-TAO  & $0.2\%$ & $9.5:1$ & $1$--$1\,\textrm{K}$ & $1$ & $1$--$4\,\textrm{KB}$ & $1$--$1\,\textrm{K}$ & $0.8$\\
\Xhline{1.5pt}
\multirow{2}{*}{TPC-C}   & {\bf New-Order} & {\bf Payment} & {\bf Delivery} & {\bf Order-Status} & {\bf Stock-Level} & {\bf Dist/WH} & {\bf WH/svr} \\
\cline{2-8}
        & $44\%$    & $44\%$    & $4\%$ & $4\%$ & $4\%$ & $10$ & $8$ \\
\Xhline{1.5pt}
\end{tabular}
\caption{Workload parameters. RO and RW mean read-only and read-write transactions, respectively. TPC-C has a scaling factor of $10$ districts per warehouse and $8$ warehouses per server.} 
\label{tbl:workload}
\end{figure*}

\paragraph{\sstx is live.} 
\sstx{}'s non-blocking execution guarantees that requests  always run to completion, 
i.e., execution never stalls (\S\ref{subsubsec:lowcost}). 
Blocking can happen only to the sending of responses due to response timing control, 
and \sstx{} avoids circular waiting with early aborts
(\S\ref{subsec:avoid-tip}). 
Thus, \sstx{} guarantees that transactions finish eventually. 

\sstx{}'s specialized read-only transaction protocol and optimization techniques such as asynchrony-aware timestamps and \sr do not affect correctness, 
because transactions are protected by the three mechanisms 
(i.e., M\textsubscript{1}, M\textsubscript{2}, and M\textsubscript{3} summarized at the beginning of this subsection) regardless of whether optimizations or the specialized protocol are used. 
 
\nps\section{Evaluation}
\label{sec:eval}
This section answers the following questions:
\begin{enumerate}[topsep=0pt,itemsep=0ex,partopsep=1ex,parsep=1ex]
    \item How well does \sstx perform, compared to common strictly serializable techniques dOCC, d2PL, and TR? 
    
    \item How well does \sstx perform, compared to state-of-the-art serializable (weaker consistency) techniques?
    
    \item How well does \sstx recover from client failures?
\end{enumerate}

\begin{figure*}[t]
\footnotesize
\centering
\renewcommand{\arraystretch}{1.5}
\begin{tabular}{@{}l cc  cc c@{}}
\Xhline{1.5pt}
{\bf Workload} & {\bf Contention} & {\bf \# shots} & {\bf Characteristics} & {\bf \sstx takeaway} \\
\hline
Facebook-TAO  & Low & $1$ & Read-dominated & Performance-optimal reads by the RO protocol \\
\hline
Google-F1  & Low & $1$ & Read-dominated & Performance-optimal reads by the RO protocol \\
\hline
TPC-C   & Medium$\rightarrow$High & Multi-shot & Write-intensive & Leverages the natural arrival order, minimizes false aborts \\
\hline
Google-WF & Low$\rightarrow$High & $1$ & Write-intensive & Leverages the natural arrival order, minimizes false aborts\\
\Xhline{1.5pt}
\end{tabular}
\caption{Facebook-TAO and Google-F1 have low contention. TPC-C and Google-WF (varying write fractions) are 
write-intensive. 
TPC-C Payment and Order-Status are multi-shot.} 
\label{tbl:takeaway}
\end{figure*}

\paragraph{Implementation.} 
We developed \sstx{} on Janus's framework~\cite{Mu:osdi2016}.
We improved the framework by making it support multi-shot transactions, optimizing its baselines, and adding more benchmarks.  
\sstx's core protocols have \tld{}3\,K lines of C++ code.
We also show the results of 
\sstxrw, a version without the read-only transaction protocol, 
i.e., all transactions are executed as read-write transactions. 

\paragraph{Baselines.} 
The evaluation includes three strict serializable baselines (dOCC, d2PL, and Janus) and two serializable baselines (MVTO and TAPIR). 
We chose d2PL and dOCC because they are the most common strictly serializable techniques. 
We chose Janus because it is the only open-source TR-based strictly serializable system we could find. 
We chose MVTO because it has the highest best-case performance among all (weaker) serializable techniques, 
presenting a performance upper bound. 
We chose TAPIR because it utilizes timestamp-based concurrency control. 

Our evaluation focuses on concurrency control 
and assumes servers never fail. 
Janus and TAPIR are unified designs of the concurrency control and replication layers, so 
we disabled their replication and only compare with their concurrency control protocols,
shown as Janus-CC and TAPIR-CC, to make the comparisons fair. 
We compare with two variants of d2PL. 
\tplnw aborts a transaction if the lock is not available. 
\tplww makes the transaction wait if it has a larger timestamp 
and aborts the lock-holding transaction otherwise. 
All baselines are fully optimized: 
we co-locate coordinators with clients (even if baselines cannot handle client failures), 
combine the execute and prepare phases for \tplnw and TAPIR-CC,  
and enable asynchronous commitment, 
i.e., the client replies to the user without waiting for the acknowledgments of 
commit messages. 

\subsection{Workloads and Experimental Setup}
We evaluate \sstx under three workloads that cover both read-dominated ``simpler'' 
transactions and many-write more ``complex'' transactions. 
Google-F1 and Facebook-TAO synthesize real-world applications 
and capture the former: they are one-shot and read-heavy. 
TPC-C has multi-shot transactions and is write-intensive, 
capturing the latter. 
We also vary write fractions in Google-F1 to further explore the latter. 
Table~\ref{tbl:workload} shows the workload parameters. 

Google-F1 parameters were published in F1~\cite{shute2013f1} and Spanner~\cite{Corbett:osdi2012}. 
Facebook-TAO parameters were published in TAO~\cite{Bronson:atc2013}. 
TPC-C's New-Order, Payment, and Delivery are read-write transactions. 
Its Order-Status and Stock-Level are read-only. 
Janus's original implementation of TPC-C is one-shot, 
so we modified it to make Payment and Order-Status multi-shot, 
to demonstrate \sstx is compatible with multi-shot transactions and 
evaluate its performance beyond one-shot 
transactions (though they are still relatively short). 

\paragraph{Experimental setting.} We use Microsoft Azure~\cite{azure}. 
Each 
machine has $4$ CPUs ($8$ cores), $16\,\textrm{GB}$ memory, and a $1\,\textrm{Gbps}$ network interface. 
We use $8$ machines as servers and $16$--$32$ machines as clients that generate open-loop requests to saturate the servers. (The open-loop clients back off when the system is overloaded to mitigate queuing delays.)
Google-F1 and Facebook-TAO have $1\,\textrm{M}$ keys, with the popular keys randomly distributed 
to balance load. 
We run $3$ trials for each test and $60$ seconds for each trial. 
Experiments are CPU-bound (i.e., handling network interrupts).

\begin{figure*}[t]
\small
\footnotesize
\centering
\begin{subfigure}[b]{0.31\linewidth}
\centering
\setlength{\abovecaptionskip}{1pt}
\setlength{\belowcaptionskip}{1pt}
  \includegraphics[width=1.0\linewidth]{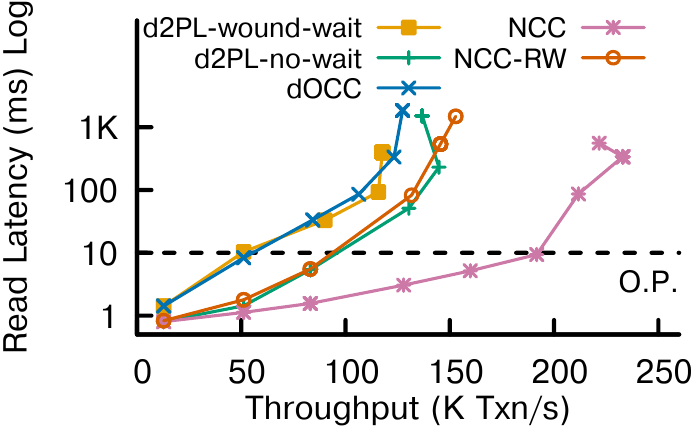}
  \caption{Google-F1 Workload}
  \label{fig:spanner-perf}
\end{subfigure}
~
\begin{subfigure}[b]{0.31\linewidth}
\centering
\setlength{\abovecaptionskip}{1pt}
\setlength{\belowcaptionskip}{1pt}
  \includegraphics[width=1.0\linewidth]{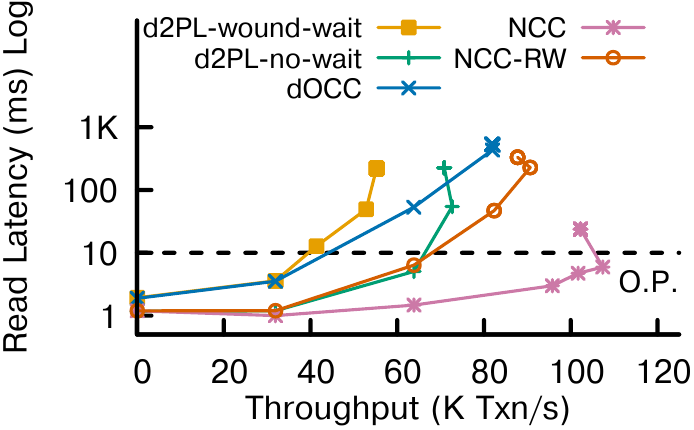}
  \caption{Facebook-TAO Workload}
  \label{fig:fb-perf}
\end{subfigure}
~
\begin{subfigure}[b]{0.31\linewidth}
\centering
\setlength{\abovecaptionskip}{1pt}
\setlength{\belowcaptionskip}{1pt}
  \includegraphics[width=1.0\linewidth]{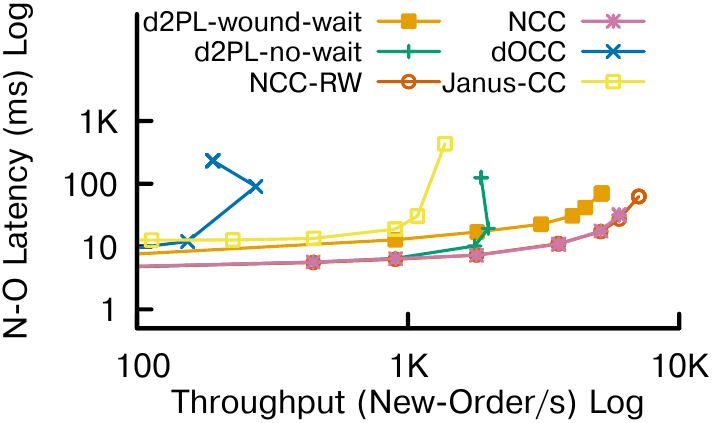}
  \caption{TPC-C Workload}
  \label{fig:tpcc-perf}
\end{subfigure}
\caption{\sstx achieves much lower latency under read-dominated workloads with its specialized read-only transaction algorithm, 50\% lower latency under write-intensive workload, and at least $\mathbf{80\%}$ higher throughput across workloads.}
\label{fig:acc-overall-perf}
\end{figure*}

\subsection{Result Overview}
\label{subsec:res-ov}
\sstx outperforms  
strictly serializable protocols dOCC, d2PL, and TR (Janus-CC) 
by $80\%$--$20\times$ higher throughput and $2$--$10\times$ lower latency under various workloads (Figure~\ref{fig:acc-overall-perf}) and 
write fractions (Figure~\ref{fig:wf}). 
\sstx outperforms and closely matches serializable systems, TAPIR-CC and MVTO, respectively (Figure~\ref{fig:ser}). 
\sstx recovers from client failures with minimal performance impact (Figure~\ref{fig:falure}). 
Please note that Figure~\ref{fig:acc-overall-perf} and Figure~\ref{fig:ser} have log-scale axes. 
Figure~\ref{tbl:takeaway} summarizes the takeaway of performance improvements. 

\subsection{Latency vs. Throughput Experiments}
\label{subsec:eval-overall}
Figure~\ref{fig:acc-overall-perf} shows \sstx{}'s overall performance is 
strictly better than the baselines, i.e., 
higher throughput with the same latency and lower latency with the same throughput. 

\paragraph{Google-F1 and Facebook-TAO.} 
Figure~\ref{fig:spanner-perf} shows the results under Google-F1. 
X-axis is the system throughput, and y-axis shows the median read latency in log scale. 
A horizontal line (O.P.) marks the operating point with reasonably low latency ($<10\,\textrm{ms}$). 
At the operating point, \sstx has a $2$--$4\times$ higher throughput than dOCC and d2PL. 
We omit the results for Janus-CC to make the graph clearer 
as we found that Janus-CC's performance is incomparable (consistently worse) with other baselines, 
because Janus-CC is designed for highly contended workloads 
by relying on heavy dependency tracking, 
which is more costly under low contention. 

\sstx has better performance because 
Google-F1 and Facebook-TAO have many naturally consistent transactions due to the prevalence of reads. 
\sstx{} enables 
low overhead 
by leveraging natural consistency. 
In particular, its read-only transaction protocol executes the 
dominating reads with the minimum costs (Figure~\ref{tbl:takeaway}).  
For instance, at the operating point, \sstx has about $99\%$ of the transactions that passed their \sg check 
and finished in one round trip. 
$99.1\%$ of the transactions did not delay their responses, 
i.e., the real-time order dependencies were already satisfied when they arrived. 
That is, $99\%$ of the transactions were finished by \sstx within a single RTT without any delays. 
For the $1\%$ of the transactions that did not pass the \sg check initially, $70\%$ of them passed the smart retry. 
Only $0.2\%$ of the transactions were aborted and retried from scratch. 
All of them were committed eventually. 

As a result, \sstx can finish most transactions 
with one round of messages (for the read-only ones) and a latency of one RTT (for both read-only and read-write) 
while dOCC and \tplww require three rounds of messages and a latency of two RTTs (asynchronous commitment saves one RTT). 
\sstx has much higher throughput than \tplnw due to its novel read-only protocol which requires 
one round of messages, while \tplnw requires two. 
The fewer messages of \sstx translate to lower latency under medium and high load due to lower queuing delay. 
\tplnw performs similar to \sstxrw because \sstxrw executes read-only transactions by following its read-write protocol. 
However, \sstxrw outperforms \tplnw under higher load because conflicts cause \tplnw to abort more frequently, while \sstxrw has fewer false aborts by leveraging the natural arrival order. 
This is more obvious in the Facebook-TAO results shown in Figure~\ref{fig:fb-perf}, because Facebook-TAO has larger 
read transactions that are more likely to conflict with writes. 
The results of Facebook-TAO show similar takeaways. 

\begin{figure*}[t]
\centering
\begin{subfigure}[b]{0.31\linewidth}
\centering
\setlength{\abovecaptionskip}{.5pt}
  \includegraphics[width=1.0\linewidth]{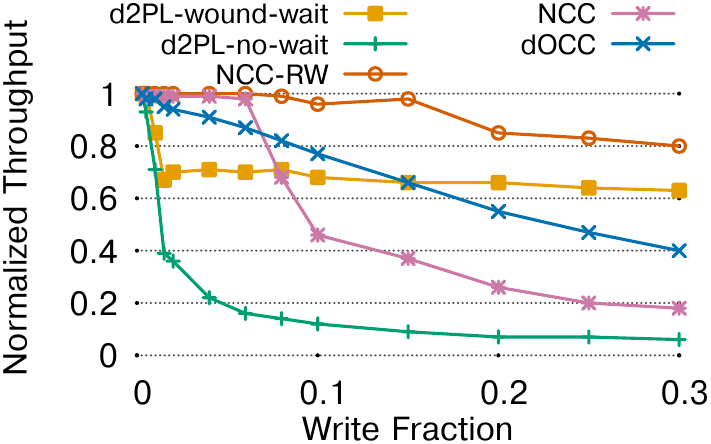}
  \caption{Varying Write Fractions}
  \label{fig:wf}
\end{subfigure}
~
\begin{subfigure}[b]{0.31\linewidth}
\centering
\setlength{\abovecaptionskip}{.5pt}
  \includegraphics[width=1.0\linewidth]{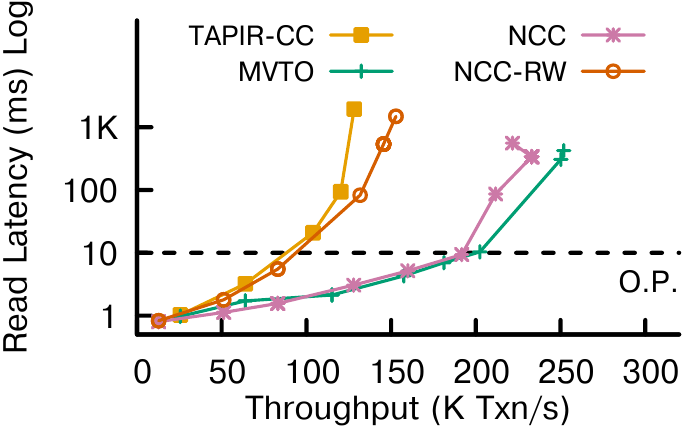}
  \caption{Weaker Serializability}
  \label{fig:ser}
\end{subfigure}
~
\begin{subfigure}[b]{0.31\linewidth}
\centering
\setlength{\abovecaptionskip}{.5pt}
  \includegraphics[width=1.0\linewidth]{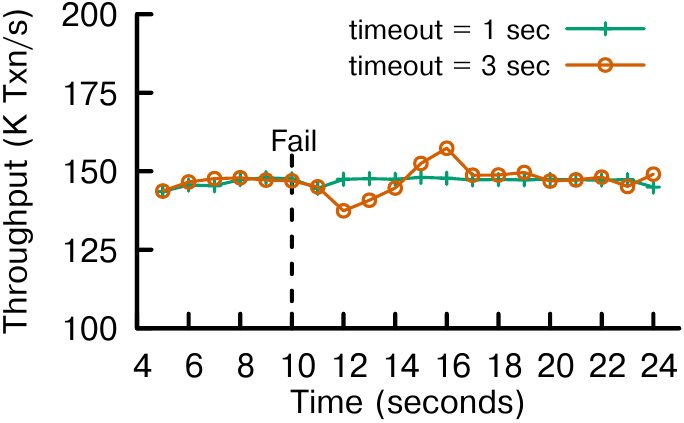}
  \caption{Failure Recovery}
  \label{fig:falure}
\end{subfigure}
\caption{\sstx{}'s performance with different write fractions (Google-WF), compared to serializable protocols (TAPIR-CC and MVTO), and under failures for the Google-F1 workload.}
\label{fig:acc-eval-details}
\end{figure*}

\paragraph{TPC-C.} 
Each experiment ran all five types of TPC-C transactions, 
and Figure~\ref{fig:tpcc-perf} shows the latency and throughput (both in log scale) of New-Order 
while the throughput of the other four types is proportional. 
\sstx and \sstxrw have \tld$20\times$ higher peak throughput with 
\tld$10\times$ lower latency compared to dOCC.  
dOCC and \tplnw have many false aborts when load increases due to conflicting writes. 
\sstx and \sstxrw can execute most naturally consistent transactions 
with low costs, even if they conflict. 
For instance, \sstxrw has more than $80\%$ of the transactions passing the \sg check and  
fewer than $10\%$ of the transactions being aborted and retried from scratch. 
\sstxrw has a $50\%$ higher peak throughput than \tplww because \sstxrw requires only two rounds of messages, while \tplww requires three. 
\sstxrw has higher peak throughput than \sstx because TPC-C has very few read-only transactions, which are also more 
likely to abort in \sstx due to conflicting writes. 
Janus-CC's performance benefits mostly come from unifying the transaction and replication layers and are less significant in a single-datacenter setting, 
especially after we made some TPC-C transactions multi-shot. 

\subsection{Additional Experiments}
\label{subsec:eval-vary}
We show more experiments with Google-F1. 
We chose Google-F1 
because 
it has both read-write and read-only transactions, while Facebook-TAO only has read-only transactions and non-transactional writes. 

\paragraph{Varying write fractions.} 
Figure~\ref{fig:wf} shows the throughput while increasing the write fraction. 
Each system is run at \tld$75\%$ load according to Figure~\ref{fig:spanner-perf}. 
The y-axis is the throughput normalized to the maximum throughput of each system during the experiment. 
The higher the write fraction, the more conflicts in the system. 
The results show that \sstxrw is most resilient to conflicts because 
\sstxrw can exploit more concurrency in those conflicting but naturally consistent transactions, i.e., \sstx has fewer aborts. 
In contrast, 
other protocols may falsely abort transactions 
due to failed validation (dOCC) or lock unavailability (d2PL variants). 
\sstx's read-only transactions are more likely to abort when writes increase 
because frequent writes cause the client to have stale knowledge of the most recently executed writes on each server; 
as a result, \sstx{} must abort the reads to avoid timestamp inversion.

\paragraph{Comparing with serializable systems.} 
Figure~\ref{fig:ser} compares \sstx with MVTO and TAPIR-CC, 
which provide serializability, under Google-F1. 
\sstx outperforms TAPIR-CC because \sstx has fewer messages with its read-only transaction 
protocol. 
MVTO and \sstx have similar performance under low and medium load because they have the same 
number of messages and RTTs. 
Under high load, MVTO outperforms \sstx when many read-only transactions in \sstx are aborted: 
MVTO never aborts reads 
because it is allowed to read stale versions, whereas 
\sstx must read the most recent version and handle timestamp inversion. 
In this sense, MVTO presents a performance upper bound for strictly serializable systems, 
and \sstx{} closely matches the upper bound. 

\paragraph{Failure recovery.} 
Figure~\ref{fig:falure} shows how well \sstxrw handles client failures under Google-F1. 
We inject failures $10$ seconds into the experiment by forcing \emph{all} clients to 
stop sending the commit messages of ongoing transactions while they continue issuing new transactions. 
Undelivered commit messages cause servers to delay the responses of later transactions 
due to response timing control, 
until the recovery mechanism is triggered after a timeout. 
We show two timeout values, $1$ and $3$ seconds. 
\sstxrw recovers quickly after failures are detected, 
thus client failures have a limited impact on throughput. 
In realistic settings, failures on one or a few clients would have a negligible impact 
because uncommitted reads do not block other reads. 
Similarly, \sstx is minimally impacted by client failures because its read-only transactions do not send commit 
messages and thus never delay later writes.

\nps\section{Related Work}
\label{sec:related}
\sstx proposes a new strictly serializable distributed protocol. 
This section places it in the context of
existing strictly serializable techniques, 
single-machine concurrency control, and techniques that provide weaker consistency. 
At a high-level, \sstx provides better performance, 
addresses a different problem setting, 
and provides stronger guarantees, compared to these categories of work, respectively.

\begin{figure*}[t]
\footnotesize
\centering
\begin{tabular}{@{}l cccccc@{}}
\midrule
{\bf System}                    & {\bf Consistency}  & {\bf Technique} & {\bf Latency (RTT)} & {\bf Lock-free} & {\bf Non-blocking} & {\bf False aborts}\\[.3ex]
\hline\\[-2.5ex]
\sstx & \textcolor{seagreen}{Strict Ser.} & NC+TS & \textcolor{seagreen}{$1$} & \textcolor{seagreen}{\textcolor{seagreen}{Yes}} & \textcolor{seagreen}{Yes} & \textcolor{seagreen}{Low} \\ 
Spanner~\cite{Corbett:osdi2012} & \textcolor{seagreen}{Strict Ser.} & d2PL+TrueTime & RO: \textcolor{seagreen}{$1$}, RW: \textcolor{red}{$2$} & RO: \textcolor{seagreen}{Yes}, RW: \textcolor{red}{No} & \textcolor{red}{No} & RO: \textcolor{seagreen}{None}, RW: \textcolor{orange}{Med} \\
d2PL-NoWait & \textcolor{seagreen}{Strict Ser.} & d2PL & \textcolor{seagreen}{$1$} & \textcolor{red}{No} & \textcolor{red}{No} & \textcolor{red}{High} \\
AOCC~\cite{adya1995efficient}  & \textcolor{seagreen}{Strict Ser.} & dOCC & \textcolor{red}{$2$} & \textcolor{seagreen}{Yes} & \textcolor{red}{No} & \textcolor{red}{High} \\[.3ex] 
Janus~\cite{Mu:osdi2016} & \textcolor{seagreen}{Strict Ser.} & TR & \textcolor{red}{$2$} & \textcolor{seagreen}{Yes} & \textcolor{red}{No} & \textcolor{seagreen}{None}\\[.3ex]
dOCC   & \textcolor{seagreen}{Strict Ser.} & dOCC & \textcolor{red}{$2$} & \textcolor{red}{No} & \textcolor{red}{No} & \textcolor{red}{High} \\[.3ex]
d2PL-WoundWait & \textcolor{seagreen}{Strict Ser.} & d2PL & \textcolor{red}{$2$} & \textcolor{red}{No} & \textcolor{red}{No} & \textcolor{orange}{Med} \\
FaRMv2~\cite{shamis2019fast}  & \textcolor{seagreen}{Strict Ser.} & dOCC & \textcolor{red}{$2$} & \textcolor{red}{No} & \textcolor{red}{No} & \textcolor{orange}{Med} \\[.3ex] 
TAPIR~\cite{zhang2018building} & \textcolor{red}{Ser.} & dOCC+TS & \textcolor{seagreen}{$1$} & \textcolor{seagreen}{Yes} & \textcolor{red}{No} & \textcolor{orange}{Med} \\[.3ex] 
DrTM~\cite{Wei:sosp2015} & \textcolor{red}{Ser.} & RO: TS, RW: d2PL & RO: \textcolor{red}{$2$}, RW: \textcolor{red}{$3$} & RO: \textcolor{seagreen}{Yes}, RW: \textcolor{red}{No} & \textcolor{red}{No} & \textcolor{orange}{Med} \\[.3ex] 
TO~\cite{bernstein1981concurrency} & \textcolor{red}{Ser.} & TS & \textcolor{seagreen}{$1$} & \textcolor{seagreen}{Yes} & \textcolor{red}{No} & \textcolor{orange}{Med} \\[.3ex] 
MVTO~\cite{reed1983implementing} & \textcolor{red}{Ser.} & TS & \textcolor{seagreen}{$1$} & \textcolor{seagreen}{Yes} & \textcolor{red}{No} & \textcolor{seagreen}{Low} \\[.3ex] 
\midrule
\end{tabular}
\caption{The consistency and best-case performance 
of representative distributed protocols for naturally consistent workloads, 
processing one-shot transactions with possible optimizations considered. 
NC means natural consistency, and TS means timestamp-based technique. 
\sstx has the lowest performance costs while providing strict serializability. 
}
\label{tbl:related}
\end{figure*}

\paragraph{General strictly serializable protocols.} 
\label{subsec:related-ss}
As discussed in Section~\ref{subsec:cc}, existing general strictly serializable protocols are 
d2PL, dOCC, TR, or their variants, 
suffering extra costs when transactions are naturally consistent. 
For instance, Spanner's read-write transactions~\cite{Corbett:osdi2012}, Sinfonia~\cite{Aguilera:sosp2007}, and Carousel~\cite{yan2018carousel} 
are variants of d2PL that must acquire locks. FaRM~\cite{Dragojevic:sosp2015}, FaRMv2~\cite{shamis2019fast},  
RIFL~\cite{Lee:sosp2015} 
are variants of dOCC that suffer extra validation costs, even if they use 
timestamp-based techniques to reduce validation aborts. 
AOCC~\cite{adya1995efficient} is a variant of dOCC and 
operates in a data-shipping environment, 
e.g., data can move from servers to client caches, 
which is different from \sstx{} which 
works in a function-shipping environment, 
i.e., data resides only on servers. 
Rococo~\cite{Mu:osdi2014} and its descendant Janus~\cite{Mu:osdi2016} reorder transactions to minimize aborts. 
Granola~\cite{Cowling:atc2012} requires an all-to-all exchange of timestamps between servers, incurring extra messages and RTTs.
Our evaluation shows that \sstx outperforms these techniques 
for real-world workloads where natural consistency is prevalent. 
When transactions are not naturally consistent, however, these techniques could outperform \sstx. 
Figure~\ref{tbl:related} summarizes performance and consistency properties of \sstx and some 
representative distributed systems.   

\paragraph{Special strictly serializable techniques.}
In addition to the general techniques discussed above, there are several interesting research directions that use specialized techniques to provide strict serializability.
Some work utilizes a centralized sequencer to 
enforce strict serializability~\cite{li2017eris, Thomson:sigmod2012, faunadb, balakrishnan2012corfu, lu2020aria, lin2019mgcrab, ren2019slog, zhou2021foundationdb}.  
Because all transactions must contact the sequencer before execution 
(e.g., Eris~\cite{li2017eris}), 
in addition to the extra latency, the sequencer can be a single point of failure and scalability bottleneck. 
Scaling out sequencers incurs extra costs,  
e.g., Calvin~\cite{Thomson:sigmod2012} 
requires all-to-all messages among sequencers for each transaction (epoch). 
Some ensure strict serializability by 
moving all data a transaction accesses to the same machine, 
e.g., LEAP~\cite{lin2016towards}. 
Some rely on program analysis and are application-dependent, e.g., the homeostasis protocol~\cite{roy2015homeostasis}. 
Some rely on extensive gossip messages for liveness, which lower throughput and increase latency, e.g., Ocean Vista~\cite{fan2019ocean} whose latency of a transaction cannot be lower than  
the gossiping delay of the slowest server even if this server is not accessed by the transaction. 
General techniques such as \sstx{} do not have the above 
limitations. 

\paragraph{Strictly serializable read-only transaction protocols.} 
To the best of our knowledge, the only existing strictly serializable read-only transaction protocol that 
has optimal \textit{best-case} performance 
is Spanner~\cite{Corbett:osdi2012}. 
Spanner ensures strict serializability by using d2PL for read-write transactions and by using synchronized clocks (TrueTime) for read-only transactions. 
TrueTime must be accurately bounded for correctness and those bounds need to be small to achieve good performance, 
which are achieved by Google's infrastructure using special hardware, e.g., GPS and atomic clocks~\cite{brewer17truetime} that are not generally available.
For instance, CockroachDB~\cite{taft2020cockroachdb}, which began as an external Spanner clone, chose not 
to support strict serializability because it does not have access to such infrastructure~\cite{cockroach_atomic_clocks}.
In contrast, 
\sstx{}'s read-only transactions achieve 
optimal best-case performance and provide strict serializability, without 
requiring synchronized clocks. 

\paragraph{Single-machine concurrency control.} 
\label{subsec:related-single}
Concurrency control for single-machine databases 
is different from the distributed setting on which this paper focuses. 
First, some techniques are not feasible in a distributed setting. 
For instance, Silo~\cite{tu2013speedy} relies on atomic instructions, and MVTL~\cite{aguilera2018locking} relies on shared lock state, which are challenging across machines. 
Second, most techniques, e.g., Silo~\cite{tu2013speedy} and TicToc~\cite{yu2016tictoc}, follow a multi-phase design and would be expensive if made distributed, 
e.g., they need distributed lock management and one round of inter-machine messages for each phase, 
which would be unnecessary costs for naturally consistent transactions. 
Their designs, however, are feasible and highly performant for the single-machine setting they target.

\paragraph{Protocols for weaker consistency.}
\label{subsec:related-mvcc}
Many systems  
trade strong consistency for better performance. 
For instance, some settle for restricted transaction APIs, e.g., read-only and/or write-only transactions~\cite{Lloyd:nsdi2013, Lloyd:sosp2011, Du:socc2013}. 
Some choose to support weaker consistency models, e.g., causal consistency and serializability~\cite{Lloyd:nsdi2013, taft2020cockroachdb, dynamodb, wei2021unifying, yu2018sundial, mahmoud2014maat, lomet2012multi, deuteronomy}. 
In contrast, \sstx provides stronger consistency and supports general transactions, 
greatly simplifying application development. 

\nps\section{Conclusion}
\label{sec:concl}
Strictly serializable datastores are advocated by recent work 
because they greatly simplify application 
development. 
This paper presents \sstx, a new design that provides 
strict serializability with minimal overhead by leveraging natural consistency 
in datacenter workloads. 
\sstx{} identifies and overcomes timestamp inversion, a fundamental correctness pitfall 
in timestamp-based concurrency control techniques.
\sstx significantly outperforms existing strictly serializable techniques 
and closely matches the performance of serializable systems.

\bibliographystyle{plain}
\nps\bibliography{ref}

\begin{thebibliography}{10}

\bibitem{adya1999weak}
Atul Adya.
\newblock {\em Weak consistency: a generalized theory and optimistic
  implementations for distributed transactions}.
\newblock PhD thesis, Massachusetts Institute of Technology, Department of
  Electrical Engineering and Computer Science, 1999.

\bibitem{adya1995efficient}
Atul Adya, Robert Gruber, Barbara Liskov, and Umesh Maheshwari.
\newblock Efficient optimistic concurrency control using loosely synchronized
  clocks.
\newblock {\em ACM SIGMOD Record}, 24(2):23--34, 1995.

\bibitem{aguilera2018locking}
Marcos~K Aguilera, Tudor David, Rachid Guerraoui, and Junxiong Wang.
\newblock Locking timestamps versus locking objects.
\newblock In {\em ACM Symposium on Principles of Distributed Computing (PODC)},
  2018.

\bibitem{Aguilera:sosp2007}
Marcos~K. Aguilera, Arif Merchant, Mehul Shah, Alistair Veitch, and Christos
  Karamanolis.
\newblock Sinfonia: A new paradigm for building scalable distributed systems.
\newblock In {\em ACM Symposium on Operating System Principles (SOSP)}, 2007.

\bibitem{infiniband}
InfiniBand~Trade Association.
\newblock Infiniband architecture specification, release 1.0, october 2000.
\newblock \url{http://www.infinibandta.org/}, 2000.

\bibitem{balakrishnan2012corfu}
Mahesh Balakrishnan, Dahlia Malkhi, Vijayan Prabhakaran, Ted Wobbler, Michael
  Wei, and John~D Davis.
\newblock {CORFU}: A shared log design for flash clusters.
\newblock In {\em USENIX Symposium on Networked Systems Design and
  Implementation (NSDI)}, 2012.

\bibitem{bernstein1981concurrency}
Philip~A Bernstein and Nathan Goodman.
\newblock Concurrency control in distributed database systems.
\newblock {\em ACM Computing Surveys (CSUR)}, 13(2):185--221, 1981.

\bibitem{google18why}
Google~Cloud Blog.
\newblock {Why you should pick strong consistency, whenever possible}.
\newblock
  \url{https://cloud.google.com/blog/products/databases/why-you-should-pick-strong-consistency-whenever-possible},
  2018.

\bibitem{brewer17truetime}
Eric Brewer.
\newblock Spanner, {TrueTime} and the {CAP} theorem.
\newblock Technical report, Google Research, 2017.

\bibitem{Bronson:atc2013}
Nathan Bronson, Zach Amsden, George Cabrera, Prasad Chakka, Peter Dimov, Hui
  Ding, Jack Ferris, Anthony Giardullo, Sachin Kulkarni, Harry Li, Mark
  Marchukov, Dmitri Petrov, Lovro Puzar, Yee~Jiun Song, and Venkat
  Venkataramani.
\newblock {TAO}: {F}acebook’s distributed data store for the social graph.
\newblock In {\em USENIX Annual Technical Conference (ATC)}, Jun 2013.

\bibitem{chen2017fast}
Haibo Chen, Rong Chen, Xingda Wei, Jiaxin Shi, Yanzhe Chen, Zhaoguo Wang, Binyu
  Zang, and Haibing Guan.
\newblock Fast in-memory transaction processing using {RDMA} and {HTM}.
\newblock {\em ACM Transactions on Computer Systems (TOCS)}, 35(1):1--37, 2017.

\bibitem{Corbett:osdi2012}
James~C. Corbett, Jeffrey Dean, Michael Epstein, Andrew Fikes, Christopher
  Frost, JJ~Furman andSanjay Ghemawat, Andrey Gubarev, Christopher Heiser,
  Peter Hochschild, Wilson Hsieh, Sebastian Kanthak, Eugene Kogan, Hongyi Li,
  Alexander Lloyd, Sergey Melnik, David Mwaura, David Nagle, Sean Quinlan,
  Rajesh Rao, Lindsay Rolig, Yasushi Saito, Michal Szymaniak, Christopher
  Taylor, Ruth Wang, and Dale Woodford.
\newblock Spanner: Google's globally-distributed database.
\newblock In {\em USENIX Symposium on Operating Systems Design and
  Implementation (OSDI)}, 2012.

\bibitem{Cowling:atc2012}
James Cowling and Barbara Liskov.
\newblock Granola: Low-overhead distributed transaction coordination.
\newblock In {\em USENIX Annual Technical Conference (ATC)}, Jun 2012.

\bibitem{dpdk}
DPDK.
\newblock {DPDK}.
\newblock \url{http://dpdk.org/}, 2020.

\bibitem{Dragojevic:sosp2015}
Aleksandar Dragojevic, Dushyanth Narayanan, Edmund~B. Nightingale, Matthew
  Renzelmann, Alex Shamis, Anirudh Badam, and Miguel Castro.
\newblock No compromises: distributed transactions with consistency,
  availability, and performance.
\newblock In {\em ACM Symposium on Operating System Principles (SOSP)}, Oct
  2015.

\bibitem{Du:socc2013}
Jiaqing Du, Sameh Elnikety, Amitabha Roy, and Willy Zwaenepoel.
\newblock Orbe: Scalable causal consistency using dependency matrices and
  physical clocks.
\newblock In {\em ACM Symposium on Cloud Computing (SoCC)}, 2013.

\bibitem{dynamodb}
Amazon DynamoDB.
\newblock {Amazon DynamoDB :: Fast and flexible {N}o{SQL} database service for
  any scale}.
\newblock \url{http://aws.amazon.com/dynamodb/}, 2021.

\bibitem{fan2019ocean}
Hua Fan and Wojciech Golab.
\newblock Ocean {V}ista: Gossip-based visibility control for speedy
  geo-distributed transactions.
\newblock {\em Proceedings of the VLDB Endowment (PVLDB)}, 12(11):1471--1484,
  2019.

\bibitem{faunadb}
FaunaDB.
\newblock {FaunaDB :: The data API for your client‑serverless applications}.
\newblock \url{https://fauna.com/}, 2021.

\bibitem{Fischer:pds1983}
Michael~J. Fischer, Nancy~A. Lynch, and Michael~S. Paterson.
\newblock Impossibility of distributed consensus with one faulty process.
\newblock In {\em Proc. Principles of Database Systems}, Mar 1983.

\bibitem{garcia1992main}
Hector Garcia-Molina and Kenneth Salem.
\newblock Main memory database systems: An overview.
\newblock {\em IEEE Transactions on knowledge and data engineering},
  4(6):509--516, 1992.

\bibitem{gifford1981information}
David~K. Gifford.
\newblock {\em Information storage in a decentralized computer system}.
\newblock PhD thesis, Stanford University, Department of Electrical
  Engineering, 1981.

\bibitem{grosvenor2015queues}
Matthew~P Grosvenor, Malte Schwarzkopf, Ionel Gog, Robert~NM Watson, Andrew~W
  Moore, Steven Hand, and Jon Crowcroft.
\newblock Queues don’t matter when you can {JUMP} them!
\newblock In {\em USENIX Symposium on Networked Systems Design and
  Implementation (NSDI)}, 2015.

\bibitem{herlihy90linearizability}
Maurice~P. Herlihy and Jeannette~M. Wing.
\newblock Linearizability: A correctness condition for concurrent objects.
\newblock {\em ACM Transactions on Programming Languages and Systems (TOPLAS)},
  12(3):463--492, 1990.

\bibitem{hstore_one_shot}
Robert Kallman, Hideaki Kimura, Jonathan Natkins, Andrew Pavlo, Alexander
  Rasin, Stanley Zdonik, Evan~PC Jones, Samuel Madden, Michael Stonebraker,
  Yang Zhang, John Hugg, and Daniel~J. Abadi.
\newblock H-store: A high-performance, distributed main memory transaction
  processing system.
\newblock {\em Proceedings of the VLDB Endowment (PVLDB)}, 1(2):1496--1499,
  2008.

\bibitem{cockroach_atomic_clocks}
Spencer Kimball and Irfan Sharif.
\newblock Living without atomic clocks.
\newblock
  \url{https://www.cockroachlabs.com/blog/living-without-atomic-clocks/}, 2021.

\bibitem{konwar2021snow}
Kishori~M Konwar, Wyatt Lloyd, Haonan Lu, and Nancy Lynch.
\newblock {SNOW} revisited: Understanding when ideal read transactions are
  possible.
\newblock In {\em IEEE International Parallel and Distributed Processing
  Symposium (IPDPS)}, 2021.

\bibitem{kraska2013mdcc}
Tim Kraska, Gene Pang, Michael~J Franklin, Samuel Madden, and Alan Fekete.
\newblock {MDCC}: Multi-data center consistency.
\newblock In {\em ACM SIGOPS European Conference on Computer Systems
  (EuroSys)}, 2013.

\bibitem{lamport78time}
Leslie Lamport.
\newblock Time, clocks, and the ordering of events in a distributed system.
\newblock {\em Communications of the ACM (CACM)}, 21(7), 1978.

\bibitem{leslie1998part}
Leslie Lamport.
\newblock The part-time parliament.
\newblock {\em ACM Transactions on Computer Systems (TOCS)}, 16(2):133--169,
  1998.

\bibitem{lamport2001paxos}
Leslie Lamport.
\newblock Paxos made simple.
\newblock {\em ACM SIGACT News}, 32(4):18--25, 2001.

\bibitem{Lee:sosp2015}
Collin Lee, Seo~Jin Park, Ankita Kejriwal, Satoshi Matsushitay, and John
  Ousterhout.
\newblock Implementing linearizability at large scale and low latency.
\newblock In {\em ACM Symposium on Operating System Principles (SOSP)}, 2015.

\bibitem{deuteronomy}
Justin Levandoski, David Lomet, Sudipta Sengupta, Ryan Stutsman, and Rui Wang.
\newblock High performance transactions in deuteronomy.
\newblock In {\em Conference on Innovative Data Systems Research (CIDR)}, 2015.

\bibitem{li2017eris}
Jialin Li, Ellis Michael, and Dan~RK Ports.
\newblock Eris: Coordination-free consistent transactions using in-network
  concurrency control.
\newblock In {\em ACM Symposium on Operating System Principles (SOSP)}, 2017.

\bibitem{li1988multiprocessor}
Kai Li and Jeffrey~F Naughton.
\newblock Multiprocessor main memory transaction processing.
\newblock In {\em Proceedings International Symposium on Databases in Parallel
  and Distributed Systems}, pages 177--178. IEEE Computer Society, 1988.

\bibitem{lin2016towards}
Qian Lin, Pengfei Chang, Gang Chen, Beng~Chin Ooi, Kian-Lee Tan, and Zhengkui
  Wang.
\newblock Towards a non-2{PC} transaction management in distributed database
  systems.
\newblock In {\em ACM International Conference on Management of Data (SIGMOD)},
  2016.

\bibitem{lin2019mgcrab}
Yu-Shan Lin, Shao-Kan Pi, Meng-Kai Liao, Ching Tsai, Aaron Elmore, and
  Shan-Hung Wu.
\newblock Mg{C}rab: Transaction crabbing for live migration in deterministic
  database systems.
\newblock {\em Proceedings of the VLDB Endowment (PVLDB)}, 12(5):597--610,
  2019.

\bibitem{Lloyd:sosp2011}
Wyatt Lloyd, Michael~J. Freedman, Michael Kaminsky, and David~G. Andersen.
\newblock Don't settle for eventual: Scalable causal consistency for wide-area
  storage with {COPS}.
\newblock In {\em ACM Symposium on Operating System Principles (SOSP)}, 2011.

\bibitem{Lloyd:nsdi2013}
Wyatt Lloyd, Michael~J. Freedman, Michael Kaminsky, and David~G. Andersen.
\newblock Stronger semantics for low-latency geo-replicated storage.
\newblock In {\em USENIX Symposium on Networked Systems Design and
  Implementation (NSDI)}, 2013.

\bibitem{lomet2012multi}
David Lomet, Alan Fekete, Rui Wang, and Peter Ward.
\newblock Multi-version concurrency via timestamp range conflict management.
\newblock In {\em IEEE International Conference on Data Engineering (ICDE)},
  2012.

\bibitem{Lu:osdi2016}
Haonan Lu, Christopher Hodsdon, Khiem Ngo, Shuai Mu, and Wyatt Lloyd.
\newblock The {{SNOW}} theorem and latency-optimal read-only transactions.
\newblock In {\em USENIX Symposium on Operating Systems Design and
  Implementation (OSDI)}, 2016.

\bibitem{lu-osdi23-ncc}
Haonan Lu, Shuai Mu, Siddhartha Sen, and Wyatt Lloyd.
\newblock {NCC}: Natural concurrency control for strictly serializable
  datastores by avoiding the timestamp-inversion pitfall.
\newblock In {\em USENIX Symposium on Operating Systems Design and
  Implementation (OSDI)}, 2023.

\bibitem{lu2020performance}
Haonan Lu, Siddhartha Sen, and Wyatt Lloyd.
\newblock Performance-optimal read-only transactions.
\newblock In {\em USENIX Symposium on Operating Systems Design and
  Implementation (OSDI)}, 2020.

\bibitem{lu:osdi20tr}
Haonan Lu, Siddhartha Sen, and Wyatt Lloyd.
\newblock Performance-optimal read-only transactions (extended version).
\newblock Technical report, TR-005-20 v1, Princeton University, Department of
  Computer Science, 2020.

\bibitem{Lu:sosp2015}
Haonan Lu, Kaushik Veeraraghavan, Philippe Ajoux, Jim Hunt, Yee~Jiun Song,
  Wendy Tobagus, Sanjeev Kumar, and Wyatt Lloyd.
\newblock Existential consistency: Measuring and understanding consistency at
  {Facebook}.
\newblock In {\em ACM Symposium on Operating System Principles (SOSP)}, Oct
  2015.

\bibitem{lu2020aria}
Yi~Lu, Xiangyao Yu, Lei Cao, and Samuel Madden.
\newblock Aria: A fast and practical deterministic {OLTP} database.
\newblock {\em Proceedings of the VLDB Endowment (PVLDB)}, 13(12):2047--2060,
  2020.

\bibitem{mahmoud2014maat}
Hatem~A Mahmoud, Vaibhav Arora, Faisal Nawab, Divyakant Agrawal, and Amr
  El~Abbadi.
\newblock {MaaT}: Effective and scalable coordination of distributed
  transactions in the cloud.
\newblock {\em Proceedings of the VLDB Endowment (PVLDB)}, 7(5):329--340, 2014.

\bibitem{azure}
Microsoft.
\newblock {Microsoft Azure :: New challenges need agile solutions. Invent with
  purpose.}
\newblock \url{https://azure.microsoft.com/en-us/}, 2020.

\bibitem{mills1992rfc1305}
David Mills.
\newblock {\em {RFC1305}: Network Time Protocol (Version 3) Specification,
  Implementation}.
\newblock RFC Editor, 1992.

\bibitem{ntp}
David~L. Mills.
\newblock Internet time synchronization: the network time protocol.
\newblock {\em IEEE Transactions on communications}, 39(10), 1991.

\bibitem{mittal2015timely}
Radhika Mittal, Vinh~The Lam, Nandita Dukkipati, Emily Blem, Hassan Wassel,
  Monia Ghobadi, Amin Vahdat, Yaogong Wang, David Wetherall, and David Zats.
\newblock Timely: {RTT}-based congestion control for the datacenter.
\newblock {\em ACM SIGCOMM Computer Communication Review}, 45(4):537--550,
  2015.

\bibitem{montazeri2018homa}
Behnam Montazeri, Yilong Li, Mohammad Alizadeh, and John Ousterhout.
\newblock Homa: A receiver-driven low-latency transport protocol using network
  priorities.
\newblock In {\em ACM Special Interest Group on Data Communication (SIGCOMM)},
  2018.

\bibitem{Mu:osdi2014}
Shuai Mu, Yang Cui, Yang Zhang, Wyatt Lloyd, and Jinyang Li.
\newblock Extracting more concurrency from distributed transactions.
\newblock In {\em USENIX Symposium on Operating Systems Design and
  Implementation (OSDI)}, 2014.

\bibitem{Mu:osdi2016}
Shuai Mu, Lamont Nelson, Wyatt Lloyd, and Jinyang Li.
\newblock Consolidating concurrency control and consensus for commits under
  conflicts.
\newblock In {\em USENIX Symposium on Operating Systems Design and
  Implementation (OSDI)}, 2016.

\bibitem{Papadimitriou79serializability}
Christos~H. Papadimitriou.
\newblock The serializability of concurrent database updates.
\newblock {\em Journal of the ACM}, 26(4), 1979.

\bibitem{raiciu2019ndp}
Costin Raiciu and Gianni Antichi.
\newblock {NDP}: Rethinking datacenter networks and stacks two years after.
\newblock {\em ACM SIGCOMM Computer Communication Review}, 49(5):112--114,
  2019.

\bibitem{reed1983implementing}
David~P Reed.
\newblock Implementing atomic actions on decentralized data.
\newblock {\em ACM Transactions on Computer Systems (TOCS)}, 1(1):3--23, 1983.

\bibitem{ren2019slog}
Kun Ren, Dennis Li, and Daniel~J Abadi.
\newblock {SLOG}: Serializable, low-latency, geo-replicated transactions.
\newblock {\em Proceedings of the VLDB Endowment (PVLDB)}, 12(11):1747--1761,
  2019.

\bibitem{roy2015homeostasis}
Sudip Roy, Lucja Kot, Gabriel Bender, Bailu Ding, Hossein Hojjat, Christoph
  Koch, Nate Foster, and Johannes Gehrke.
\newblock The homeostasis protocol: Avoiding transaction coordination through
  program analysis.
\newblock In {\em ACM International Conference on Management of Data (SIGMOD)},
  2015.

\bibitem{shamis2019fast}
Alex Shamis, Matthew Renzelmann, Stanko Novakovic, Georgios Chatzopoulos,
  Aleksandar Dragojevi{\'c}, Dushyanth Narayanan, and Miguel Castro.
\newblock Fast general distributed transactions with opacity.
\newblock In {\em ACM International Conference on Management of Data (SIGMOD)},
  2019.

\bibitem{shute2013f1}
Jeff Shute, Radek Vingralek, Bart Samwel, Ben Handy, Chad Whipkey, Eric
  Rollins, Mircea Oancea, Kyle Littlefield, David Menestrina, Stephan Ellner,
  John Cieslewicz, Ian Rae, Traian Stancescu, and Himani Apte.
\newblock F1: A distributed {SQL} database that scales.
\newblock {\em Proceedings of the VLDB Endowment (PVLDB)}, 2013.

\bibitem{stonebraker2018end}
Michael Stonebraker, Samuel Madden, Daniel~J Abadi, Stavros Harizopoulos, Nabil
  Hachem, and Pat Helland.
\newblock The end of an architectural era: It's time for a complete rewrite.
\newblock In {\em Making Databases Work: the Pragmatic Wisdom of Michael
  Stonebraker}, pages 463--489. Association for Computing Machinery and Morgan
  \& Claypool, 2018.

\bibitem{taft2020cockroachdb}
Rebecca Taft, Irfan Sharif, Andrei Matei, Nathan VanBenschoten, Jordan Lewis,
  Tobias Grieger, Kai Niemi, Andy Woods, Anne Birzin, Raphael Poss, Paul
  Bardea, Amruta Ranade, Ben Darnell, Bram Gruneir, Justin Jaffray, Lucy Zhang,
  and Peter Mattis.
\newblock Cockroach{DB}: The resilient geo-distributed {SQL} database.
\newblock In {\em ACM International Conference on Management of Data (SIGMOD)},
  2020.

\bibitem{Thomson:sigmod2012}
Alexander Thomson, Thaddeus Diamond, Shu-Chun Weng, Kun Ren, Philip Shao, and
  Daniel~J. Abadi.
\newblock Calvin: Fast distributed transactions for partitioned database
  systems.
\newblock In {\em ACM International Conference on Management of Data (SIGMOD)},
  2012.

\bibitem{tpcc}
TPC.
\newblock {TPC-C}: An on-line transaction processing benchmark.
\newblock \url{http://www.tpc.org/tpcc/}, 2020.

\bibitem{tu2013speedy}
Stephen Tu, Wenting Zheng, Eddie Kohler, Barbara Liskov, and Samuel Madden.
\newblock Speedy transactions in multicore in-memory databases.
\newblock In {\em ACM Symposium on Operating System Principles (SOSP)}, 2013.

\bibitem{wei2021unifying}
Xingda Wei, Rong Chen, Haibo Chen, Zhaoguo Wang, Zhenhan Gong, and Binyu Zang.
\newblock Unifying timestamp with transaction ordering for {MVCC} with
  decentralized scalar timestamp.
\newblock In {\em USENIX Symposium on Networked Systems Design and
  Implementation (NSDI)}, 2021.

\bibitem{Wei:sosp2015}
Xingda Wei, Jiaxin Shi, Yanzhe Chen, Rong Chen, and Haibo Chen.
\newblock Fast in-memory transaction processing using {RDMA} and {HTM}.
\newblock In {\em ACM Symposium on Operating System Principles (SOSP)}, 2015.

\bibitem{whitney1997high}
Arthur Whitney, Dennis Shasha, and Stevan Apter.
\newblock High volume transaction processing without concurrency control, two
  phase commit, {SQL} or {C++}, 1997.

\bibitem{yan2018carousel}
Xinan Yan, Linguan Yang, Hongbo Zhang, Xiayue~Charles Lin, Bernard Wong,
  Kenneth Salem, and Tim Brecht.
\newblock Carousel: Low-latency transaction processing for globally-distributed
  data.
\newblock In {\em ACM International Conference on Management of Data (SIGMOD)},
  2018.

\bibitem{yu2016tictoc}
Xiangyao Yu, Andrew Pavlo, Daniel Sanchez, and Srinivas Devadas.
\newblock {TicToc}: Time traveling optimistic concurrency control.
\newblock In {\em ACM International Conference on Management of Data (SIGMOD)},
  2016.

\bibitem{yu2018sundial}
Xiangyao Yu, Yu~Xia, Andrew Pavlo, Daniel Sanchez, Larry Rudolph, and Srinivas
  Devadas.
\newblock Sundial: Harmonizing concurrency control and caching in a distributed
  {OLTP} database management system.
\newblock {\em Proceedings of the VLDB Endowment (PVLDB)}, 11(10):1289--1302,
  2018.

\bibitem{Zhang:sosp2015}
Irene Zhang, Naveen~Kr. Sharma, Adriana Szekeres, Arvind Krishnamurthy, and Dan
  R.~K. Ports.
\newblock Building consistent transactions with inconsistent replication.
\newblock In {\em ACM Symposium on Operating System Principles (SOSP)}, 2015.

\bibitem{zhang2018building}
Irene Zhang, Naveen~Kr Sharma, Adriana Szekeres, Arvind Krishnamurthy, and
  Dan~RK Ports.
\newblock Building consistent transactions with inconsistent replication.
\newblock {\em ACM Transactions on Computer Systems (TOCS)}, 35(4):1--37, 2018.

\bibitem{zhou2021foundationdb}
Jingyu Zhou, Meng Xu, Alexander Shraer, Bala Namasivayam, Alex Miller, Evan
  Tschannen, Steve Atherton, Andrew~J Beamon, Rusty Sears, John Leach, Dave
  Rosenthal, Xin Dong, Will Wilson, Ben Collins, David Scherer, Alec Grieser,
  Young Liu, Alvin Moore, Bhaskar Muppana, Xiaoge Su, and Vishesh Yadav.
\newblock Foundation{DB}: A distributed unbundled transactional key value
  store.
\newblock In {\em ACM International Conference on Management of Data (SIGMOD)},
  2021.

\end{thebibliography}

\appendix
\nps\section{A Discussion on TAPIR}
\label{app:tapir}
TAPIR's novelty was a unified design of transaction and replication layers, 
and was not in the transaction layer in particular. 
Whether providing strict serializability or weaker serializability does not 
affect their contributions in the unified design. 
Our work identifies the cases where it is not strictly serializable and helps 
TAPIR users and the works inspired by TAPIR avoid the correctness violations caused by 
misassuming the consistency model. 
TAPIR's authors confirmed our findings, 
i.e., TAPIR is not strictly serializable and is subject to \inversion{}. 

\subsection{TAPIR's Read-Write Transactions}

Here we summarize how TAPIR works, which is based on the TOCS version~\cite{zhang2018building}.
TAPIR is built on a multi-versioned data store.
A transaction in TAPIR goes through three phases: execution, validation, and commit.
In the execution phase, the transaction reads the data items in the read set from the servers,
which return the most recently committed versions of each data item and their timestamps.
Writes are buffered locally on the client.
In the validation phase, TAPIR chooses a proposed timestamp for this transaction, which is the
combination of the machine's current physical time and the client identifier.
The proposed timestamp is sent together with the validation messages to servers.
When a server receives a validation message, it validates the transaction with the proposed timestamp.
If all servers successfully validate the transaction, TAPIR will commit it; otherwise, it will abort the transaction. 

\begin{figure}[t]
\centering
\frame{\includegraphics[width=0.85\columnwidth]{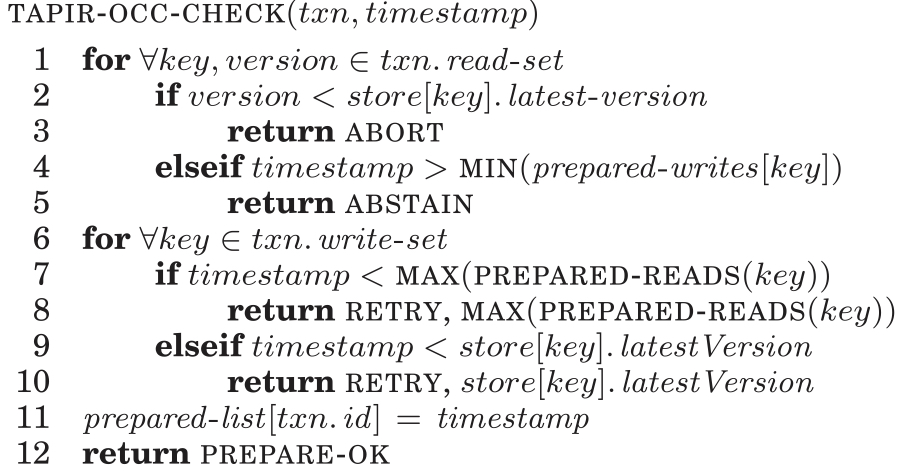}}
\caption{TAPIR's timestamp-based validation, a screenshot of Figure 9 in the TOCS version~\cite{zhang2018building}.}
\label{fig:tapir}
\end{figure}

Figure~\ref{fig:tapir} shows the pseudocode of TAPIR's validation mechanism, which is a screenshot of Figure 9 in the paper. 
TAPIR fails a read validation if the data item has newer versions than the one returned in the execution phase (lines 2, 3). 
TAPIR's validation is lock-free. 
TAPIR accepts a write validation if both conditions are satisfied: 
1. the write's proposed timestamp is greater than the maximum 
proposed timestamp of prepared-but-uncommitted reads on this data item (skipping lines 7, 8); and 
2. the write's proposed timestamp is greater than the timestamp of the most recent version (skipping lines 9, 10). 

In addition, TAPIR implements a third check: a write is aborted if there are any pending (prepared) conflicting writes.
This third check was informed by the authors during our conversations and was not included 
in the paper.

\subsection{A Counterexample}
\label{app:tapir-ex}
As shown in Figure~\ref{fig:tapir-ex}, 
There are four transactions $\textit{tx}_1$, $\textit{tx}_2$, $\textit{tx}_3$, and $\textit{tx}_4$ by three different users. 
$\textit{tx}_1$ and $\textit{tx}_4$ are from user $U_1$; 
$\textit{tx}_2$ is from user $U_2$; 
and $\textit{tx}_3$ is from user $U_3$. 
($U_2$ and $U_3$ are not shown in the figure.) 
$W$ and $R$ denote a write and a read request, respectively. 
$U_1$ issues transaction $\textit{tx}_1$ which 
reads $A$ and 
then reads $B$, and (if $B$ returns $0$, then) resets $C$ to $0$. 
Initially $A$, $B$, and $C$ each store value $0$ at timestamp $0$.
Meanwhile, 
$U_2$ issues $\textit{tx}_2$ that updates $B$ to $1$, with $W_2$. 
After $U_2$ is notified that $\textit{tx}_2$ is committed, 
$U_3$ issues $\textit{tx}_3$ which updates $C$ to $2$, with $W_3$. 
Because $\textit{tx}_3$ is issued after $\textit{tx}_2$ was finished, $\textit{tx}_3$ is in real-time after $\textit{tx}_2$, i.e., $\textit{tx}_2 \rt \textit{tx}_3$. 
$\textit{tx}_2$ and $\textit{tx}_3$ are blind writes only for simplicity, and 
the counterexample still holds if they are read-modify-writes.  
After $\textit{tx}_1$ is committed, $U_1$ issues a read-only transaction $\textit{tx}_4$ that reads $B$ and $C$. 
The events of TAPIR processing these transactions happen in the following order: 
\begin{itemize} 
\item[$1.$] $U_1$ starts $\textit{tx}_1$. 
\item[$2.$] $\textit{tx}_1$ reads $A$, reads $B$, and buffers $W_{1C}$ locally. 
\item[$3.$] $\textit{tx}_1$ chooses a proposed timestamp, $t_1 = 8$. 
\item[$4.$] $\textit{tx}_1$ validates $R_{1A}$ at $A$.
\item[$5.$] $\textit{tx}_1$ validates $R_{1B}$ at $B$. 
\item[$6.$] $U_2$ starts $\textit{tx}_2$. 
\item[$7.$] $\textit{tx}_2$ sends $W_{2B}$ to $B$ and validates $W_{2B}$ with a proposed timestamp, $t_2 = 10$. 
\item[$8.$] $\textit{tx}_2$ is committed and $U_2$ finishes $\textit{tx}_2$. 
\item[$9.$] $U_3$ starts $\textit{tx}_3$. 
\item[$10.$] $\textit{tx}_3$ sends $W_{3C}$ to $C$ and validates $W_{3C}$ with a proposed timestamp, $t_3 = 5$. 
\item[$11.$] $\textit{tx}_3$ is committed and $U_3$ finishes $W_3$. 
\item[$12.$] $\textit{tx}_1$ sends $W_{1C}$ to $C$ and validates $W_{1C}$. 
\item[$13.$] $\textit{tx}_1$ is committed and $U_1$ finishes $\textit{tx}_1$. 
\item[$14.$] $U_1$ starts $\textit{tx}_4$. 
\item[$15.$] $\textit{tx}_4$ reads $B$ and $C$ and returns $B=1$ and $C=0$. 
\item[$16.$] $\textit{tx}_4$ validates $B=1$ and $C=0$ at $B$ and $C$ with a proposed timestamp, $t_4 = 20$. 
\item[$17.$] $\textit{tx}_4$ is committed and returns $B=1$ and $C=0$ to $U_1$.
\end{itemize}

All $\textit{tx}_1$, $\textit{tx}_2$, $\textit{tx}_3$, and $\textit{tx}_4$ are able to pass TAPIR's validation and commit. 
$\textit{tx}_1$'s read validation on $A$ succeeds because there are no other transactions accessing $A$. 
$\textit{tx}_1$'s read validation on $B$ succeeds because $\textit{tx}_2$ has not started yet when $\textit{tx}_1$'s validation message arrives at $B$. 
$\textit{tx}_2$ is successfully validated and committed because both validation conditions (lines 7--12 in Figure~\ref{fig:tapir}) and the new assumption are met. 
Specifically, 
lines 7 and 8 are skipped because $\textit{tx}_2$ has a greater timestamp ($t_2=10$) than 
the greatest prepared read timestamp ($t_1=8$); 
lines 9 and 10 are skipped because $\textit{tx}_2$'s timestamp is larger than that of 
the latest version which is $0$; 
the third check is met because there are no prepared writes at $B$.
$\textit{tx}_3$ is successfully validated and committed because when $W_{3C}$ arrives at $C$, there are no other transactions. 
$\textit{tx}_1$'s write on $C$ succeeds because both validation conditions are met 
($t_1 > t_3$, so lines 9 and 10 are skipped; no reads happened at $C$, so lines 7 and 8 
are skipped), and the third check is met because $W_3$ has been committed when 
$W_1$ arrives and thus there are no conflicting prepared writes.  
Finally, $\textit{tx}_4$'s reads are successfully validated and returned to the user because 
there are no transactions ongoing on $B$ and $C$ when $\textit{tx}_4$ is executed and validated. 

\begin{figure}[t]
\centering
\includegraphics[width=0.85\columnwidth]{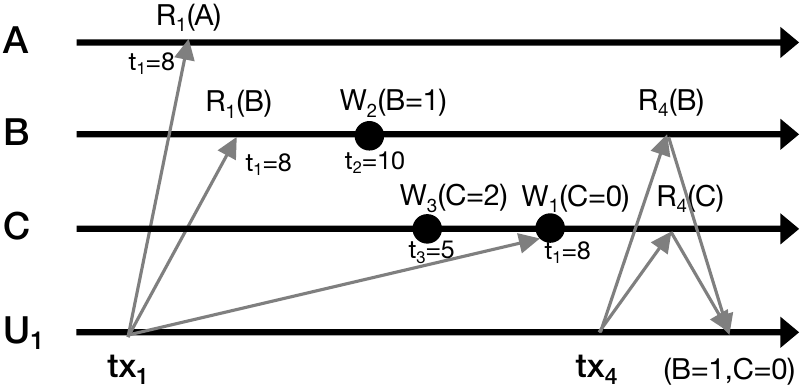}
\caption{An execution of four transactions that may 
be accepted by TAPIR but violate strict serializability.}
\label{fig:tapir-ex}
\end{figure}

Strict serializability requires that 
there exists a total order among transactions, and the total order must respect the real-time ordering. 
Because $\textit{tx}_4$ begins in real-time after $\textit{tx}_1$, $\textit{tx}_2$, and $\textit{tx}_3$ end, it must be ordered last. This leaves 6 possible total orders, which are listed below along with an explanation of why they are not a legal strictly serializable order:\\

\noindent
\begin{tabular}{@{}|l|l|@{}}
\hline
\textbf{Total order} & \\
\hline
$\textit{tx}_1$, $\textit{tx}_2$, $\textit{tx}_3$ &
\multirow{2}{.65\columnwidth}{Not legal because $\textit{tx}_4$ observes $\textit{tx}_1$'s write to C, not $\textit{tx}_3$'s} \\
$\textit{tx}_1$, $\textit{tx}_3$, $\textit{tx}_2$ & \\
\hline
$\textit{tx}_2$, $\textit{tx}_1$, $\textit{tx}_3$ & \multirow{3}{.65\columnwidth}{Not legal because $\textit{tx}_4$ observes $\textit{tx}_1$'s write to C, not $\textit{tx}_3$'s, and because $\textit{tx}_1$ observes the initial value of $B$, not $\textit{tx}_2$'s} \\
&\\
&\\
\hline
$\textit{tx}_2$, $\textit{tx}_3$, $\textit{tx}_1$ &  \multirow{2}{.65\columnwidth}{Not legal because $\textit{tx}_1$ observes the initial value of $B$, not $\textit{tx}_2$'s} \\
$\textit{tx}_3$, $\textit{tx}_2$, $\textit{tx}_1$ & \\
\hline
\multirow{2}{*}{$\textit{tx}_3$, $\textit{tx}_1$, $\textit{tx}_2$} &
\multirow{2}{.65\columnwidth}{Not strictly serializable because $\textit{tx}_2$ finishes in real-time before $\textit{tx}_3$ starts}\\
 & \\
\hline
\end{tabular}\\

This counter-example shows that TAPIR admits executions that cannot form a total order that respects the real-time order.
Therefore, TAPIR's read-write transaction protocol is not strictly serializable. 
\nps\section{A Discussion on DrTM}
\label{app:drtm}
DrTM's novelty was the first system that combined HTM and RDMA for distributed transaction processing. 
Our work identifies the cases where it is not strictly serializable and helps 
DrTM users and the works inspired by DrTM avoid the correctness violations caused by 
misassuming the consistency model. 
DrTM's authors 
confirmed that our findings are correct, i.e., DrTM's read-only transactions are subject to 
the \inversion{} pitfall. 

\subsection{DrTM's Transaction Protocol}
The discussion is based on DrTM's original paper published in SOSP~\cite{Wei:sosp2015} 
and its extended version published in TOCS~\cite{chen2017fast}.
\paragraph{Read-write transactions.}
DrTM's read-write transaction protocol is a variant of two-phase locking (2PL), 
which keeps write (exclusive) locks and replaces read (shared) locks with leases. 
A transaction first attempts to acquire write lock and/or read leases on remote machines using 
RDMA compare-and-set requests. 
We omit how write locks are managed because it is the same as that in standard 2PL. 
We focus on how read leases are managed. 

Figure~\ref{fig:drtm-rw} shows the pseudocode for the high-level structure of a distributed 
read-write transaction 
in DrTM, which is a screenshot of Figure 3 in the SOSP version. 
Before a transaction starts, the issuing server generates a timestamp, \textit{end\_time}, 
which is the server's current physical time plus a duration ($1\,\textrm{ms}$ for read-only 
and $0.4\,\textrm{ms}$ for other transactions). 
Servers' physical clocks are periodically synchronized using the precision time protocol. 
\textit{end\_time} is included in remote read requests of this transaction and specifies the 
time when the read leases will expire (if the reads are granted access to the records). 
A lease is considered expired if the server's current clock value is greater than the lease's 
\textit{end\_time} (with some \textit{DELTA} for clock skew). 
On the remote server, a read request is granted access if the data record is not write-locked. 
If there is an existing lease, its \textit{end\_time} is extended to the request's \textit{end\_time} if the latter is greater. 
The read request is then granted access. The \textit{end\_time} of the lease is also returned. 
To commit a transaction, the issuing server checks if write locks are successfully acquired and all read leases are still valid. 
The issuing server \textit{assumes} read leases are still valid if 
its current local time is smaller than the minimum \textit{end\_time} in all read responses (with some \textit{DELTA} for clock skew). 

That is,
a write is rejected only if the data record either is write-locked or 
has an active (unexpired) read lease; 
a read is rejected only if the record is write-locked. 
A read-write transaction is aborted only if either any read/write request was rejected 
or the issuing server's clock value is greater than \textit{end\_time} upon commit. 

\begin{figure}[t]
\centering
\includegraphics[width=0.85\columnwidth]{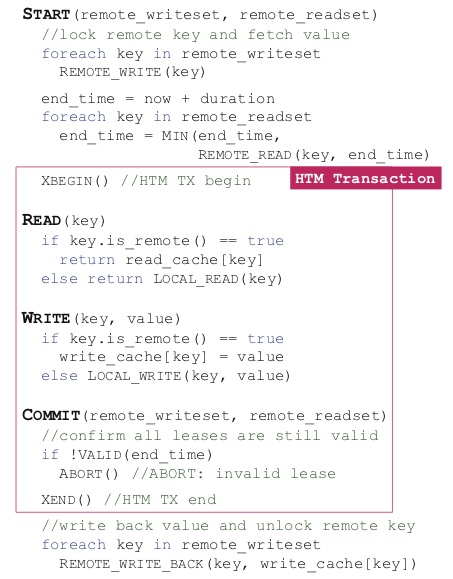}
\caption{DrTM's read-write transaction protocol, a screenshot of Figure 3 in the SOSP paper~\cite{Wei:sosp2015}.}
\label{fig:drtm-rw}
\end{figure}

\paragraph{Read-only transactions.}
A read-only transaction only involves leases since there are no writes in it. 
Figure~\ref{fig:drtm-ro} is a screenshot of the read-only transaction protocol (Figure 8 in the SOSP paper). 
The read-only transaction protocol is equivalent to the read-write transaction 
protocol with its write part taken out. 
Then, leases (timestamps) are used to order read-only transactions, i.e., they are 
ordered by \textit{end\_time}. 
A read-only transaction is aborted only if either any of its read was rejected 
or the issuing server's clock value is greater than \textit{end\_time} upon commit. 

\subsection{A Counterexample}
Figure~\ref{fig:drtm-ex} shows a counterexample where the read-only 
transaction may return system state that is not strictly serializable. 
There are three transactions $\textit{tx}_1$, $\textit{tx}_2$, and $\textit{tx}_3$, 
from users $U_1$, $U_2$, and $U_3$, respectively. 
All transactions are distributed, and we omit local HTM transactions. 
$U_1$ is the DrTM server that starts $\textit{tx}_1$, and we omit showing 
$U_2$ and $U_3$ in the figure. 
Assume each server stores one data item for simplicity. 
Both $A$ and $B$ have an initial value of $0$. 
$\textit{tx}_1$ is a read-only transaction reading both servers $A$ and $B$. 
(We combine RDMA-CAS and RDMA-READ as one message for simplicity.) 
$\textit{tx}_2$ and $\textit{tx}_3$ are single-key write transactions, having write requests, 
$W_2$ and $W_3$, respectively. 
We make $\textit{tx}_2$ and $\textit{tx}_3$ blind writes only for simplicity, and 
the counterexample still holds if they are read-modify-writes. 
After $\textit{tx}_2$ updates $A$ to $1$ and responds to $U_2$, $\textit{tx}_3$ is started 
by $U_3$ and updates $B$ to $2$, so $\textit{tx}_3$ is after $\textit{tx}_2$ in real time, 
i.e., $\textit{tx}_2 \rt \textit{tx}_3$. 
$\textit{tx}_1$ is concurrent with both $\textit{tx}_2$ and $\textit{tx}_3$, 
e.g., $R_1(A)$ arrives before $\textit{tx}_2$, and $R_1(B)$ arrives after $\textit{tx}_3$. 
\textit{Clk} is the server's physical clock time. 
$ET$ is the \textit{end\_time} of a read lease. 
In this example, we use small integers for a simpler representation of clock values.  
The events of DrTM processing these transactions happen in the following order: 

\begin{itemize} 
\item[$1.$] $U_1$ starts $\textit{tx}_1$, at its clock time $1$, and sets $end\_time$ $8$. 
\item[$2.$] $R_1(A)$ arrives at $A$, at $A$'s local time $7$, and starts a read lease on $A$ with an \textit{end\_time} of $8$.  
\item[$3.$] $\textit{tx}_2$ arrives at $A$, at $A$'s local time $9$, and is granted the write lock. 
\item[$4.$] $\textit{tx}_2$ commits and releases the write lock on $A$. 
\item[$5.$] $\textit{tx}_3$ is started and arrives at $B$, and is granted the write lock. 
\item[$6.$] $\textit{tx}_3$ commits and releases the write lock on $B$. 
\item[$7.$] $R_1(B)$ arrives at $B$, at $B$'s local time $5$, and starts a read lease on $B$ with an \textit{end\_time} of $8$.  
\item[$8.$] $U_1$ commits $\textit{tx}_1$ at its local time $7$, and returns $A=0$ and $B=2$. 
\end{itemize}

\begin{figure}[t]
\centering
\includegraphics[width=0.85\columnwidth]{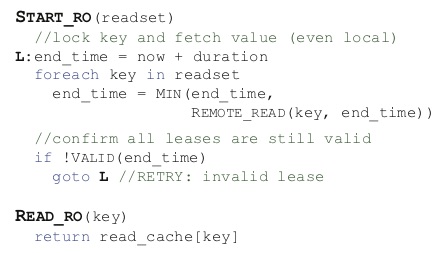}
\caption{DrTM's ro transaction protocol, screenshot of Fig 8 in the SOSP paper~\cite{Wei:sosp2015}.}
\label{fig:drtm-ro}
\end{figure}

All transactions in the execution are able to commit in DrTM. 
$R_1(A)$ is able to hold a lease on $A$ because there is no active 
lease or lock on $A$ when it arrives and because $A$'s local time is $7$, smaller than the lease's \textit{end\_time} $8$. 
$\textit{tx}_2$ is able to acquire the write lock because there is no write lock on 
$A$ and the read lease of $R_1(A)$ has expired at time $9$. 
$\textit{tx}_3$ is able to acquire the write lock on $B$ because there is no 
active lease or lock on $B$ when it arrives. 
$R_1(B)$ can hold a lease on $B$ because there is no write lock, e.g., 
$\textit{tx}_3$ has released it, and $B$'s local time is $5$, smaller than the 
\textit{end\_time} $8$. 
$\textit{tx}_1$ is committed by $U_1$ because both reads hold the leases and 
because $U_1$ ``believes'' the read leases are still valid at the commit 
time $7$, which is smaller than the \textit{end\_time} $8$. 

Strict serializability requires there is a total order of transactions and the total order respects the real-time order. 
For instance, because $\textit{tx}_2$ is before $\textit{tx}_3$ in real time, a total order 
is legal only if $\textit{tx}_2$ is before $\textit{tx}_3$ in the total order. 
DrTM's execution constructs a total order of 
$\textit{tx}_3 \ex \textit{tx}_1 \ex \textit{tx}_2$, because 
$\textit{tx}_1$ observes $\textit{tx}_3$'s state but not $\textit{tx}_2$'s. 
However, this total order violates the real-time order: $\textit{tx}_2 \rt \textit{tx}_3$. 
Therefore, DrTM's read-only transactions are not strictly serializable. 
DrTM's authors believe that synchronized clocks such as TrueTime must be used (and the protocol needs update accordingly to work with TrueTime) to make the clock skew \textit{DELTA} \textit{relative to the ground truth of time} to be accurately captured in order to possibly avoid the above counterexample from happening. 

\begin{figure}[t]
\centering
\includegraphics[width=0.85\columnwidth]{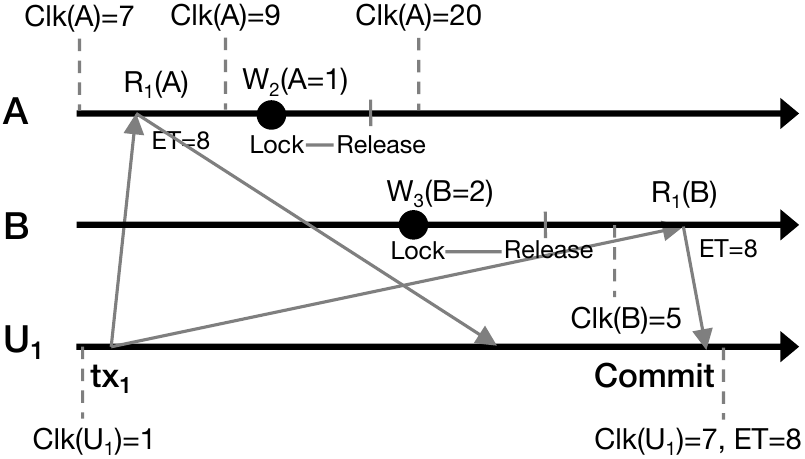}
\caption{An execution of three transactions that would 
be accepted by DrTM but violate strict serializability.}
\label{fig:drtm-ex}
\end{figure}

\nps\section{Supporting Complex Transaction Logic}
\label{app:detail}
\sstx{}
supports complex logic, 
e.g., a transaction accesses the same key multiple times via  
read-modify-writes and repeated reads/writes, 
by treating its requests to the same key as a single logical operation. 

\paragraph{Read-modify-write.} 
If a read-modify-write has its read and write requests executed consecutively 
(i.e., they are not intersected by other writes), 
then only the write response is checked by the \sg, 
treating read-modify-write as one logical request; 
otherwise, it is aborted if there are intersecting writes, 
e.g., when the most recent version has a $t_w$ greater than that returned by the read of 
this read-modify-write. 
The response of the read is not sent (thus the modify-write part will not start) until 
it is allowed by response timing control. 
After the modify-write part is executed, its write response is inserted right after the read response of the same read-modify-write, and sent back to the client immediately. 
This is safe because its read response has satisfied all real-time dependencies and was safely sent by response timing control, so its own write response is also safe to send together with the read response. 
In the commit phase, the client sends commit/abort messages for both read and write requests in the read-modify-write to the servers. 

\paragraph{Read-after-write.}
When the transaction writes to a key and then later reads the same key, 
\sstx{} does not send a request to the server for the read. 
Instead, the client feeds the read with the value of its own earlier write to the key, and  buffers the read 
locally. 
For read-after-write, the \sg{} only checks the write response (together with the responses of the other requests of this transaction) for consistency, and the result of the read (which is the value of its own write) will be returned at the end to the user when the transaction is committed.

\paragraph{Repeated reads and repeated writes.}
When the transaction reads/writes the same key multiple times, these requests are executed, checked (by the \sg{} and \sr{}), and committed/aborted normally, following  \sstx{}'s protocol. 

\nps\section{Details on Failure Handling}
\label{app:fh}
This section presents the details of how \sstx handles client failures. 
In particular, we must handle the scenario where the client has made a commit (or 
abort) decision for a transaction and sent its results to the user, 
but failed before (during) sending the commit/abort messages to the servers. 
\sstx{} must make servers arrive at the same commit/abort decision as the client would if the client did not fail, i.e., making the final state consistent with what was returned to the user. 
Figure~\ref{pc:failure} shows the pseudocode. 

\begin{figure}[t]
\hrule\ 
\footnotesize
\begin{Verbatim}[commandchars=\\\{\}, numbers=left, numbersep=1ex, xleftmargin=2ex]
record   # \CM{record kept by each tx}
txs[tid] # \CM{ongoing transactions indexed by tx id}

\KW{function} handle_failure(tx):
  \KW{if} tx.record.status == uncleared; \KW{then}
    tx.record.status = aborted # \CM{avoid deadlocks}
    AsyncCommitOrAbort(tx, ``aborted'') # \CM{Algorithm 5.2}
    \KW{return}
  \KW{if} tx.record.coord == this_svr; \KW{then} # \CM{coordinator}
    status = resolve_status(tx)
  \KW{else} # \CM{cohort}
    status = resolve_status_coord(tx.tid)
  \KW{if} status == uncleared; \KW{then}
    \KW{return} # \CM{do nothing}
  \KW{else} # \CM{finalize this tx}
    AsyncCommitOrAbort(tx, status) # \CM{Algorithm 5.2}

\KW{function} resolve_status(tx):
  \KW{if} tx.record.status != cleared; \KW{then}
    \KW{return} tx.record.status
  records = \{\}
  \KW{for} svr \KW{in} tx.record.cohorts; \KW{do} # \CM{RPC to cohorts}
    record = svr.get_record(tx.tid)
    records.add(record)
  \KW{for} record \KW{in} records; \KW{do}
    \KW{if} record.status == committed; \KW{return} committed
    \KW{if} record.status == aborted; \KW{return} aborted
    \KW{if} record.status == uncleared; \KW{return} uncleared
  # \CM{all parts in tx are cleared} 
  # \CM{reconstruct client decision} 
  # \CM{with record's \{(t\textsubscript{w}, t\textsubscript{r})\}} field
  \KW{invoke} lines 8--16 in Algorithm 5.1
  
\KW{function} get_record(tid): 
  \KW{if} txs.find(tid) == nil; \KW{then} 
    record = \{\}; record.status = aborted
    \KW{return} record
  tx = txs[tid] 
  \KW{return} tx.record
  
\KW{function} resolve_status_coord(tid):
  \KW{if} txs.find(tid) == nil; \KW{then} \KW{return} aborted
  tx = txs[tid]
  \KW{return} resolve_status(tx) 
\end{Verbatim}
\hrule\
\vspace{1ex}
\caption{\sstx{} client failure handling.}
\label{pc:failure}
\end{figure}

Each transaction keeps a \textit{record} that consists of 
four fields: \textit{status}, \{($t_w$, $t_r$)\}, \textit{coord}, and \textit{cohorts} (line 1). 
\textit{status} stores the current status of the transaction: 
\textit{uncleared}, \textit{cleared}, \textit{committed}, or \textit{aborted}. 
When a transaction arrives at the server for the first time, 
its \textit{record} is created with \textit{status} initially being uncleared. 
\textit{status} is changed to cleared when 
the server has received the last-shot requests of the transaction (a flag, \textit{IS\_LAST\_SHOT}, in the requests indicates the last shot) \textit{and} 
the responses of this transaction are ready to be sent by response timing control. 
That is, \textit{status} is cleared when a transaction has finished its logic and is 
ready to commit or abort. 
\textit{status} is changed from cleared to either committed or aborted 
when the server has received the commit 
messages. 
\textit{status} is changed from uncleared to aborted when this transaction early aborts to avoid indefinite waits.  
When \textit{status} is aborted, future requests of this transaction are ignored.

\{($t_w$, $t_r$)\} stores the set of ($t_w$, $t_r$) pairs in request responses. 
\textit{coord} stores the identity of the backup coordinator, which is chosen by the client and is typically one of the servers the transaction accesses.  
\textit{cohorts} stores the identities of the other participant servers this 
transaction accesses. 
The identity of the backup coordinator is piggybacked on each request. 
The identities of cohorts are sent to the backup coordinator in the last shot. 
The client will not start the commit phase, e.g., the \sg check, until it has received the responses of the last-shot requests/messages. 

After a \textit{record} is created for a transaction, 
a function, \textit{handle\_failure}, is 
registered and a timer starts. 
\textit{handle\_failure} is invoked every time a timeout $t$ fires. 
The system suspects the client failed if the server has not received the commit messages within time $t$.  
The timer is canceled and \textit{handle\_failure} is deleted when the transaction is committed or aborted. 
Our evaluation shows two values for $t$, $1$ and $3$ seconds. 

If \textit{status} is still uncleared when \textit{handle\_failure} is invoked, \textit{status} is set to aborted 
and we abort this transaction. (\textit{handle\_failure} will not be invoked if this 
transaction has been committed or aborted.) 
It is safe to unilaterally abort an uncleared transaction because the client must have 
not received all responses and thus must have not made a commit/abort decision.  
(For details, see the proof in Appendix~\ref{app:failure}.) 
If the server is the backup coordinator, then it resolves the final decision of this 
transaction by contacting the cohorts with \textit{get\_record} RPCs (lines 9--10, 22--24). 
When a cohort receives a \textit{get\_record} request, it returns what it knows about this 
transaction. 
For instance, if the cohort does not have a \textit{record} for this transaction, which means 
it has not received the requests, it returns \textit{status} $=$ \textit{aborted} (lines 35--37). 
Otherwise, the cohort returns the \textit{record} of this transaction. 

After receiving all the responses of \textit{get\_record}, the coordinator 
resolves the final decision. 
If any cohort replies committed, it commits this transaction (line 26). 
If any cohort replies aborted, it aborts this transaction (line 27). 
If any cohort replies uncleared, it does nothing in this round of \textit{handle\_failure} and restarts the timer 
(lines 28, 13, 14). 
If all cohorts reply cleared, which means all participant servers are waiting for 
the commit messages from the possibly-failed client, the coordinator 
reconstructs the client decision deterministically based on the same inputs used in transaction execution. 
That is, the coordinator invokes the same \sg{} check with the same ($t_w,\,t_r$) pairs and \sr{} if necessary (lines 29--32). 
Note that, if the client did a smart retry and it succeeded, then a re-issued smart retry by the coordinator must succeed; if the earlier smart retry failed, then 
the re-issued one must also fail. 
This is because the timestamps are in the past and have already been properly updated by the original smart retried.
Due to deterministic computation, the coordinator must arrive at the same commit/abort decision for this transaction as the client would have if it did not fail. 

When a cohort receives a message that invokes \textit{handle\_failure} to a transaction, the transaction is aborted if it is still uncleared. 
Otherwise, the cohort sends a \textit{resolve\_status\_coord} request to the backup coordinator (line 12). 
When the coordinator receives the request, it goes through the \textit{resolve\_status} logic, which could then send \textit{get\_record} requests to all cohorts for resolving the decision 
(lines 41--44). 
When the cohort receives the response of \textit{resolve\_status\_coord}, it commits or 
aborts the transaction accordingly (lines 15, 16). 
If the response is uncleared, then it does nothing in this round of \textit{handle\_failure} (lines 13, 14). 

\sstx{}'s failure handling guarantees that the backup coordinator always makes the same decision for a transaction as the client would have if the client did not fail. 
We present a formal proof of correctness in Appendix~\ref{app:failure}.

\nps\section{Proof of Correctness}
\label{app:proof}
This section provides a correctness proof of \sstx. 
We prove that \sstx is safe, i.e., it guarantees strict serializability, 
and is live, i.e., transactions eventually terminate. 
We also prove that the failure handling mechanism is safe and live. 
We first state some preliminaries (\S\ref{app:proof-pre}) and then provide a proof summary (\S\ref{app:proof-sum}) followed by the main proof (\S\ref{app:sstx}
and \S\ref{app:failure}). 

\subsection{Preliminaries}
\label{app:proof-pre}
\paragraph{System model.} We adopt a system model similar to that used in FLP~\cite{Fischer:pds1983}. A distributed datastore is modeled as a set of $N$ processes, $P_1, P_2, \ldots, P_n$. We denote the set of front-end client machines by $c$ and the server machines that store the data by $S$. Processes communicate by sending and receiving messages.

Transaction processing is modeled as I/O automata: each process (either a server or a client) is an automaton that implements a deterministic state machine. 
Upon receiving a message, a process does local computation based on its current state and the message input, then moves to the next state and sends out one or more output messages. For instance, upon receiving a transaction request (input message), the server executes this request against its current state by either reading or modifying the data and associated metadata according to the transaction protocol (local computation and state transition), and then sends the response to the client (output message).

For simplicity and clarity, we consider local computation, e.g., executing a transaction request, as one atomic step even though they could consist of a sequence of sub-state transitions in general, e.g., memory updates, disk updates, etc.

\paragraph{Assumptions.} Our system model assumes the following:
\begin{itemize} 
\item[$A_1.$] The system network is reliable: messages are eventually delivered, guaranteed by most distributed datastores.
\item[$A_2.$] Each process (e.g., a server or a client) has a physical clock that is not perfectly synchronized. Clocks can be loosely synchronized, e.g., via NTP~\cite{ntp}. 
\item[$A_3.$] Both clients and servers can fail. Failure handling was discussed in Section~\ref{subsec:failure} and 
Appendix~\ref{app:fh}.
\end{itemize}

\subsection{Proof Summary}
\label{app:proof-sum}
The proof consists of two parts. 
The first part (\S\ref{app:sstx}) proves that \sstx ensures strict serializability and is live. 
Specifically, 
\reflem{totalorder} proves that committing all parts of each transaction at the same timestamp, i.e., the one intersecting all returned ($t_w,\,t_r$) pairs, guarantees a total order. 
In particular, this total order is aligned with the numerical order of timestamps. 
\reflem{realtime} proves that \sstx satisfies the real-time order by proving that if a transaction is included in a consistent snapshot, then all transactions that in real-time precede it are included in the snapshot. 
\reflem{liveness} proves liveness by demonstrating that early aborts ensure it is impossible for responses to circularly wait on each other during RTC.  
The second part (\S\ref{app:failure}) proves that client failure handling is safe and live by 
proving three invariants enforced by the protocol.  

\subsection{Correctness of \sstx}
\label{app:sstx}
\begin{definition}[execution order] 
We define the execution order between transactions (requests) as follows. 
Two requests from different transactions $\textit{req}_1 \ex \textit{req}_2$ if any of the following happens: 
$\textit{req}_1$ creates some data version $v_i$ and $\textit{req}_2$ reads $v_i$; 
$\textit{req}_1$ reads some data version $v_j$ and $\textit{req}_2$ creates $v$'s next version that is after $v_j$; 
or $\textit{req}_1$ creates some data version $v_k$ and $\textit{req}_2$ creates $v$'s next version that is after $v_k$. 
Two transactions $\textit{tx}_1 \ex \textit{tx}_2$ if there exist $\textit{req}_1$ and $\textit{req}_2$ from $\textit{tx}_1$ and $\textit{tx}_2$, respectively, such that $\textit{req}_1 \ex \textit{req}_2$.  
Execution order is transitive in that if $\textit{tx}_1 \ex \textit{tx}_2$ and $\textit{tx}_2 \ex \textit{tx}_3$, then $\textit{tx}_1 \EX \textit{tx}_3$. 
Execution order represents the order in which transactions take effect. 
\label{def:order}
\end{definition}

\begin{definition}[total order]
We construct a Directed Serialization Graph (DSG)~\cite{adya1999weak}, which 
presents the execution order between transactions. 
Vertices in a DSG are all committed transactions in the system. 
A directed execution edge is drawn from transaction $\textit{tx}_1$ to $\textit{tx}_2$ if $\textit{tx}_1 \ex \textit{tx}_2$. 
Thus, we use the same symbol, $\ex$, to denote both the execution order between transactions and the execution edge in a DSG. 
Then, $\textit{tx}_1 \EX \textit{tx}_2$ denotes a directed path in the DGS from $\textit{tx}_1$ to $\textit{tx}_2$. 
All committed transactions construct a total order if and only if their DSG is acyclic, 
i.e., the following invariant holds:  
\label{def:totalorder}
\end{definition}

\setword{\textbf{Inv\_total}}{total}{: $\forall \textit{tx}_1,\,\textit{tx}_2\,(\textit{tx}_1 \EX \textit{tx}_2 \implies \neg (\textit{tx}_2 \EX \textit{tx}_1))$}

\begin{definition}[real-time order]
The lifetime of a transaction \textit{tx} starts with its invocation, \INV{tx}, i.e., when the client sends out the transaction requests to servers, and ends with its response, \RESP{tx}, i.e., when the user has received the results of the transaction from the client. 
Two transactions, $\textit{tx}_1$ and $\textit{tx}_2$, have a real-time ordering relationship if $\textit{tx}_2$'s invocation happens in real-time after $\textit{tx}_1$'s response, i.e., $\textit{tx}_1 \rt \textit{tx}_2$. 
We construct Real-time Serialization Graphs (RSGs)~\cite{adya1999weak}, 
which augment DSGs by introducing a real-time edge from transaction $\textit{tx}_1$ to $\textit{tx}_2$ if $\textit{tx}_1 \rt \textit{tx}_2$. 
We use $\rt$ to denote both the real-time ordering between transactions
and the real-time edge in an RSG. 
The real-time ordering requirement of strict serializability requires that 
if $\textit{tx}_1 \rt \textit{tx}_2$, then $\textit{tx}_1$ must be ordered before $\textit{tx}_2$ in the total order. 
That is, the following invariant holds: 
\label{def:realtime}
\end{definition}

\setword{\textbf{Inv\_realtime}}{realtime}{: $\forall \textit{tx}_1,\,\textit{tx}_2\,(\textit{tx}_1 \rt \textit{tx}_2 \implies \neg (\textit{tx}_2 \EX \textit{tx}_1))$}

\begin{definition}[strict serializability] A concurrency control protocol guarantees strict serializability if and only if there exists a total order among all transactions (Definition~\ref{def:totalorder}), and the total order respects the real-time order (Definition~\ref{def:realtime}). 
That is, for any execution allowed by the protocol, both~\ref{total} and~\ref{realtime} hold. 
\label{def:ss}
\end{definition}

\begin{lemma}
\label{thm:totalorder}
Natural Concurrency Control (\sstx{}) guarantees a total order among transactions (\ref{total}).
\end{lemma}
\proofstart
We prove by contradiction. 
Assume that there exist committed transactions $\textit{tx}_1$ and $\textit{tx}_n$ 
such that $\textit{tx}_1 \EX \textit{tx}_n$ and $\textit{tx}_n \EX \textit{tx}_1$. 
Then, there exist a chain of committed transactions such that 
$\textit{tx}_1 \ex \textit{tx}_2 \ex \ldots \ex \textit{tx}_{n-1} \ex \textit{tx}_n$, 
by $\textit{tx}_1 \EX \textit{tx}_n$ and Definition~\ref{def:order}. 
Let \{($t_{wi},\,t_{ri}$)\} be the set of returned timestamp pairs for each transaction $i$, where $i \in [1, n]$. 
Because all transactions in the chain are committed, 
\{($t_{wi},\,t_{ri}$)\} must overlap for each transaction $\textit{tx}_i$,  by \sstx{}'s commit protocol, 
i.e., the \sg{} and \sr{} logic. 
Let $t_{ci}$ be the greatest $t_{w}$ that intersects \{($t_{wi},\,t_{ri}$)\}. 
$t_{ci}$ is the commit timestamp (or synchronization point) of $\textit{tx}_i$. 

Consider $\textit{tx}_1 \ex \textit{tx}_2$. 
Then, there exist $\textit{req}_1$ and $\textit{req}_2$ in $\textit{tx}_1$ and $\textit{tx}_2$, 
respectively, such that $\textit{req}_1 \ex \textit{req}_2$, by Definition~\ref{def:order}. 
Let ($t_{w1},\,t_{r1}$) and ($t_{w2},\,t_{r2}$) be the returned timestamp pairs of $\textit{req}_1$ and $\textit{req}_2$, respectively. 
There are three cases, by Definition~\ref{def:order} and \sstx{}'s protocol. 

Case 1: $\textit{req}_1$ is a write and $\textit{req}_2$ is a write, then $t_{w1} = t_{r1} < t_{w2} = t_{r2}$. 
Case 2: $\textit{req}_1$ is a read and $\textit{req}_2$ is a write, then $t_{w1} \le t_{r1} < t_{w2} = t_{r2}$. 
Case 3: $\textit{req}_1$ is a write and $\textit{req}_2$ is a read, then $t_{w1} = t_{r1} = t_{w2} \le t_{r2}$.
Thus, we can derive $t_{w1} \le t_{r1} \le t_{w2} \le t_{r2}$. 
Because $t_{w1} \le t_{c1} \le t_{r1}$ and $t_{w2} \le t_{c2} \le t_{r2}$ where $t_{c1}$ and $t_{c2}$ are the commit timestamps of $\textit{tx}_1$ and $\textit{tx}_2$, respectively, 
we have $t_{c1} \le t_{c2}$. 
By applying the same argument to the rest of the chain: 
$\textit{tx}_2 \ex \ldots \ex \textit{tx}_{n-1} \ex \textit{tx}_n$, 
we can derive $t_{c1} \le t_{c2} \le \ldots \le t_{cn}$, which implies $t_{c1} \le t_{cn}$. 

Similarly, by applying the same argument to the assumption $\textit{tx}_n \EX \textit{tx}_1$, 
we can derive $t_{cn} \le t_{c1}$. 
This implies that $t_{c1} = t_{cn}$. 
Given $t_{c1} = t_{cn}$ and $\textit{tx}_1 \EX \textit{tx}_n$, there must exist $\textit{req}_1$ and $\textit{req}_n$ in $\textit{tx}_1$ and $\textit{tx}_n$, respectively, such that $\textit{req}_1 \ex \textit{req}_n$, 
by \sstx{}'s protocol, and $t_{w1} = t_{wn} = t_{c1} = t_{cn}$, by the definition of $t_c$ and~\ref{def:order}.  
Then, $\textit{req}_1$ and $\textit{req}_n$ must be a write and a read, respectively, 
by \sstx{}'s protocol (because only reading a write can result in the latter request, i.e., the read, returning the same timestamp pairs as the former request, i.e., the write). 
Similarly, because $t_{cn} = t_{c1}$ and $\textit{tx}_n \EX \textit{tx}_1$, we can derive that there must exist $\textit{req}_n'$ and $\textit{req}_1'$ in $\textit{tx}_n$ and $\textit{tx}_1$, respectively, such that $t_{wn}' = t_{w1}' = t_{c1} = t_{cn}$, where $\textit{req}_n'$ and $\textit{req}_1'$ are a write and a read, respectively.  
Then, $t_{w1} = t_{wn}'$, which is impossible because $t_{w1}$ and $t_{wn}'$ are the returned $t_w$ of write requests $\textit{req}_1$ and $\textit{req}_n'$, respectively, and writes from different transactions must have distinct timestamps, i.e., \sstx{}'s timestamps are unique.
Because all four cases lead to contradictions, we have proved the lemma is true.
\proofend

\begin{lemma}
\label{thm:realtime}
\sstx guarantees the real-time ordering between transactions (\ref{realtime}).
\end{lemma}
\proofstart
We prove by contradiction. Let $\textit{tx}_1$ and $\textit{tx}_2$ be two arbitrary transactions. 
Let $t_1$ be the time when \RESP{$\textit{tx}_1$} happens and $t_2$ be the time when \INV{$\textit{tx}_2$} happens. 
Without loss of generality, we assume $\textit{tx}_1$ is in real-time before $\textit{tx}_2$, i.e., $t_1 < t_2$ (Definition~\ref{def:realtime}). 
Assume for the sake of contradiction that $\textit{tx}_2 \EX \textit{tx}_1$. 
Then, there exists  
a set of $k$ transactions $\textit{tx}_{i1},\,\textit{tx}_{i2},\,\ldots,\, \textit{tx}_{ik}$, such that $\textit{tx}_2 \ex \textit{tx}_{i1} \ex \textit{tx}_{i2} \ex \ldots \ex \textit{tx}_{ik} \ex \textit{tx}_1$, for $k \geq 0$ (Definition~\ref{def:order}). 
Then, for each transaction $\textit{tx}$, there exist two requests $\textit{req}$ and $\textit{req}'$ (not necessarily distinct), such that 
$\textit{req}_2' \ex \textit{req}_{i1}$, $\textit{req}_{i1}' \ex \textit{req}_{i2}$, $\ldots$ , $\textit{req}_{ik}' \ex \textit{req}_1$, by Definition~\ref{def:order}. 
For each request, let $t_u$ and $t_v$ be the physical (real) time when the server sends its response and receives a commit message, respectively. 
Then, $t_{v2}' < t_{u(i1)}$, $t_{v(i1)}' < t_{u(i2)}$, $\ldots$ , $t_{v(ik)}' < t_{u1}$, 
by response timing control. 
Specifically, $\textit{tx}_2$ and $\textit{tx}_{i1}$ must access at least one common data item 
by the definition of ``$\ex$'', and thus RTC ensures that $\textit{tx}_{i1}$ will not be responded until $\textit{tx}_2$ is committed/aborted. 
Because the server can receive the commit messages only after responses of all requests have been received by the client, $t_u < t_v$ for any pair of requests, $\textit{req}$ and $\textit{req}'$, in the same transaction. 
Thus, $t_{u2} < t_{v2}' < t_{u(i1)} < t_{v(i1)}' < t_{u(i2)} < \ldots < t_{v(ik)}' < t_{u1}$. 
Thus, $t_{u2} < t_{u1}$. 
Because $t_2 < t_{u2}$ and $t_{u1} < t_1$, we can derive $t_2 < t_{u2} < t_{u1} < t_1$, 
which implies $t_2 < t_1$,  
contradicting the assumption, $t_1 < t_2$. 
Therefore, 
the lemma is true. 
\proofend

\begin{lemma}
\label{thm:liveness}
\sstx transactions do not prevent each other from making progress, i.e., they eventually commit or abort.
\end{lemma}
\proofstart
Waiting happens only to server responses due to response timing control, i.e., the sending of a response waits until its real-time dependencies are met (\S\ref{subsec:avoid-tip}). 
Because response queues are kept one per key, circular waiting is only possible when a transaction $\textit{tx}_1$ is executed before $\textit{tx}_2$ on one key but after $\textit{tx}_2$ on another key and they have write conflicts on both servers, i.e., interleaved read-write or write-write conflicts. 
However, we prove that at least one transaction must abort and thus break the cycle of waiting. 

Let $\textit{tx}_1$ and $\textit{tx}_2$ be interleaved read-write transactions that both access keys $A$ and $B$. 
Without loss of generality, let $\textit{tx}_1$ arrive before $\textit{tx}_2$ on key $A$ and after $\textit{tx}_2$ on key $B$. 
Suppose there is at least one write on each server by these two transactions, so circular waiting is possible.  
Because $\textit{tx}_1$ and $\textit{tx}_2$ are interleaved and have write conflicts, 
they must be both waiting for each other to be committed/aborted. 
Let $t_1$ and $t_2$ be the pre-assigned timestamps of $\textit{tx}_1$ and 
$\textit{tx}_2$, respectively. 
Without loss of generality, assume $t_1 < t_2$ (timestamps are all unique). 
Then, $\textit{tx}_1$ must abort early on $B$ 
because $\textit{tx}_1$ arrives later and has a smaller timestamp, while $\textit{tx}_2$ has not been committed/aborted, 
by \sstx{}'s protocol (specifically, the logic of early aborts). 
Therefore, circular waiting is not possible in \sstx. 
Because read-only transactions never block later writes, 
i.e., they do not insert responses into the response queues or send commit messages, 
they do not cause other writes to wait. 
Moreover, read-only transactions eventually commit/abort because their responses only wait for the preceding writes, which will eventually commit/abort, as argued above. 
\proofend

\vspace{1em}
\noindent\textbf{\textit{Proof of \sstx.}}
We have proved that Natural Concurrency Control (\sstx) guarantees strict serializability by proving that it ensures a total order (\reflem{totalorder}) and the total order respects the real-time order (\reflem{realtime}). 
We also have proved that \sstx is live by proving that it is impossible for transactions to depend on each other while no transaction can proceed to commit/abort (\reflem{liveness}). 
\proofend

\subsection{Correctness of Client Failure Handling} 
\label{app:failure}
This section proves the correctness of failure handling by proving that 
the following invariants always hold for any transaction \textit{tx} when \textit{handle\_failure} is triggered, i.e., 
\textit{tx}'s client either fails or is very slow. 
(Unspecified line numbers referenced in the proof are those in Figure~\ref{pc:failure}.)  
\begin{itemize} 
\item[$I_1.$] If the client has not made a commit/abort decision for \textit{tx}, 
i.e., \textit{tx} has not entered its commit phase, 
then \textit{tx} will eventually abort, i.e., there are no dangling transactions on servers.  
\item[$I_2.$] If the client has decided to commit/abort \textit{tx}, then \textit{tx} will always be committed or aborted by the servers following the client's decision. 
\item[$I_3.$] Failure handling is live, i.e., \textit{tx} does not prevent concurrent transactions issued by either healthy or failed clients from making progress. That is, transactions will eventually commit or abort when failures are present. 
\end{itemize}

\begin{lemma}
\label{thm:fh-1}
Invariant $I_1$ always holds.
\end{lemma}
\proofstart
Let $C$ be the client that fails and \textit{tx} be an arbitrary transaction by $C$ when $C$ fails. 
Assume \textit{tx} has not entered the commit phase. 
Then, either \textit{tx} has not finished executing its logic or $C$ has not received all of the server responses.  
That is, at least one of the following cases must occur, corresponding to each of the three dependencies. 

\textbf{Case 1:} There exists at least one server $S$, such that \textit{tx} accesses $S$ but $S$ has not 
received \textit{tx}'s last-shot request. Then, \textit{tx}'s \textit{record} does not exist on $S$ if \textit{tx} is a one-shot transaction, 
or \textit{tx}'s \textit{record} must have uncleared \textit{status} if \textit{tx} is a multi-shot 
transaction because \textit{status} is cleared only after a transaction has received requests for all shots. 
Then, in either case, \textit{tx} will abort (lines 5, 35, 42). 

\textbf{Case 2:} There exists at least one read $r$ in \textit{tx}, such that $r$ waits for version $v$, which was created by write $w$ of another transaction, to be committed (Dependency~\ref{dep1}). 
Then, \textit{tx}'s \textit{status} must be uncleared because $r$'s response is not ready to be sent. 
Then, $S$ will abort \textit{tx} (lines 5--8). 
By a similar argument to Case 1, the decision of aborting \textit{tx} will be passed to \textit{tx}'s cohorts. 

\textbf{Case 3:} There exists at least one write $w$ in \textit{tx}, such that $w$ is waiting for preceding 
transactions to commit/abort (Dependency~\ref{dep2} and~\ref{dep3}). 
Similar to the argument in Case 2, \textit{tx} will abort, and the decision will be 
passed to the cohorts. 

According to the above three cases, covering the three dependencies, \textit{tx} will eventually abort on all its participant servers. 
Therefore, the lemma is true.
\proofend

\begin{lemma}
\label{thm:fh-2}
Invariant $I_2$ always holds.
\end{lemma}
\proofstart
Let $C$ be the client that fails and \textit{tx} be an arbitrary transaction by $C$ when $C$ fails. 
Assume that \textit{tx} has entered the commit phase, and $C$ has made a commit/abort decision for \textit{tx}. Then, 
there are two cases.

\textbf{Case 1:} $C$ made a commit decision for \textit{tx}. 
Because \textit{tx} is committed, \textit{tx} must pass the \sg check. 
Then, all servers involved in \textit{tx} must have sent the responses, which include ($t_w,\,t_r$) pairs. 
Let $S$ be an arbitrary server that is involved in \textit{tx}. 
If $S$ had received \textit{tx}'s commit/abort message before $C$ failed, then \textit{handle\_failure} is not triggered on $S$ since failure timeout is cancelled upon receiving the commit/abort message. 
\textit{tx}'s \textit{status}, which is committed, will be passed to the other participant servers when $S$ is queried via \textit{get\_record} and 
\textit{resolve\_status\_coord}. 
If $S$ did not receive \textit{tx}'s commit/abort message before $C$ failed, then 
\textit{tx}'s \textit{status} on $S$ must be cleared. 
If $S$ is the coordinator, then \textit{get\_record} will be triggered to query the cohorts. 
If any of the cohorts return committed, then $S$ will commit \textit{tx}. 
Otherwise, if all cohorts reply cleared, then $S$ will invoke the same \sg{} check logic to 
reconstruct the decision for \textit{tx} (lines 29--32). 
Because the set of input and the \sg{} logic are the same as those used by $C$, $S$ will 
derive the same decision for \textit{tx} as $C$, which is committed, by the fact that the computation is deterministic.

\textbf{Case 2:} $C$ made an aborted decision for \textit{tx}.  
By the same arguments in Case 1, we can derive that $S$ will make the same 
decision for \textit{tx} as the client $C$, which is to abort. 

Because $S$ makes the same commit or abort decision for \textit{tx} as the client would make, 
the lemma is true.
\proofend

\begin{lemma}
\label{thm:fh-3}
Invariant $I_3$ always holds.
\end{lemma}
\proofstart
We have proved that \sstx is live without failures (\reflem{liveness}). 
When client failures are present, an affected transaction \textit{tx} could block future transactions during \textit{handle\_failure}. 
That is, \textit{tx} cannot be committed/aborted until \textit{handle\_failure} is done, 
and future transactions cannot be committed/aborted before \textit{tx} commits/aborts due to response timing control. 
We will prove that such situation is impossible. 
There are four cases while resolving \textit{tx}'s \textit{status} via \textit{handle\_failure}.

\textbf{Case 1:} \textit{tx}'s \textit{status} is committed on at least one participant server (coordinator or cohort). 
Then, this \textit{status} will be passed to the other servers via \textit{resolve\_status\_coord} and \textit{get\_record} (lines 34, 41). 
Then, \textit{tx} will be committed on all servers, thus unblocking future transactions. 

\textbf{Case 2:} \textit{tx}'s \textit{status} is aborted on at least one participant server (coordinator or cohort). 
Then, \textit{tx} will abort on all servers by the same argument in Case 1, 
thus unblocking future transactions. 

\textbf{Case 3:} \textit{tx}'s \textit{status} is cleared on all servers. 
Then, the coordinator will reconstruct the client decision for \textit{tx} by invoking the \sg check, which will 
make \textit{tx}'s final \textit{status} be either committed or aborted (lines 29--32). This reconstructed decision will be 
passed to the cohorts, which will commit/abort \textit{tx} accordingly 
and unblock future transactions.

\textbf{Case 4:} \textit{tx}'s \textit{status} is uncleared on at least one server. 
Then, \textit{tx}'s \textit{status} will eventually become aborted on this server when \textit{handle\_failure} is invoked (lines 5--8). 
This final \textit{status} will be passed to the other servers via either \textit{resolve\_status\_coord} or \textit{get\_record}. 
Therefore, \textit{tx} will be eventually aborted on all participant servers, 
thus unblocking future transactions. 

Because \textit{tx}'s \textit{status} is eventually resolved in all cases, \textit{tx} will not prevent other transactions from 
making progress. 
\proofend

\vspace{1em}
\noindent\textbf{\textit{Proof of failure handling.}}
Because all three invariants always hold, \sstx{}'s client failure handling  is safe and live. 
\proofend
}{
}

\end{document}